\documentclass[preprintnumbers, floatfix, letterpaper, twocolumn,aps,prd, nofootinbib]{revtex4-1}
\pdfoutput=1
\usepackage{bm,graphicx,dcolumn,epstopdf,epsf, latexsym,mathbbol, amssymb,amsmath,color,slashed, mathrsfs,mathcomp,simplewick}
\usepackage[center]{subfigure}
\usepackage{multirow}
\usepackage{makecell}
\usepackage{float}
\usepackage{booktabs}
\usepackage[colorlinks,linkcolor=blue,citecolor=blue,urlcolor=blue]{hyperref}
\begin{document}
%\allowdisplaybreaks
%%%%%%%%%%%%%%%%%%%%%%%%
 \newcommand{\bq}{\begin{equation}} 
 \newcommand{\eq}{\end{equation}}
 \newcommand{\bqn}{\begin{eqnarray}}
 \newcommand{\eqn}{\end{eqnarray}}
 \newcommand{\nb}{\nonumber}
 \newcommand{\lb}{\label}
 \newcommand{\f}{\frac}
 \newcommand{\p}{\partial}
%%%%%%%%%%%%%%%%%%%%%%%%%
\newcommand{\PRL}{Phys. Rev. Lett.}
\newcommand{\PLB}{Phys. Lett. B}
\newcommand{\PRD}{Phys. Rev. D}
\newcommand{\CQG}{Class. Quantum Grav.}
\newcommand{\JCAP}{J. Cosmol. Astropart. Phys.}
\newcommand{\JHEP}{J. High. Energy. Phys.}
\newcommand{\Doi}{https://doi.org}
\newcommand{\arXiv}{https://arxiv.org/abs}
 %%%%%%%%%%%%%%%%%%%%%%%%
\title{Pre-inflationary dynamics of Starobinsky inflation and its generization in Loop Quantum Brans-Dicke Cosmology}

%\title{Pre-inflationary dynamics of Starobinsky inflation and its generalization in Loop Quantum Brans-Dicke Cosmology}
%
\author{Wei-Jian Jin${}^{a}$}

\author{Yongge Ma${}^{b}$}

\author{Tao Zhu${}^{a}$}
\email{zhut05@zjut.edu.cn}

%
%\author{Klaus Kirsten${}^{c}$}
%\email{klaus$\_$kirsten@baylor.edu} 
%
%\author{Gerald Cleaver${}^{d}$}
%\email{gerald$\_$cleaver@baylor.edu} 
%
%\author{Qin Sheng${}^{c}$}
%\email{qin$\_$sheng@baylor.edu} 

\affiliation{${}^{a}$ Institute for Theoretical Physics $\&$ Cosmology, Zhejiang University of Technology, Hangzhou, 310032, China\\
${}^{b}$Department of Physics, Beijing Normal University, Beijing 100875, China}
%\\
%${}^{c}$ GCAP-CASPER, Mathematics Department, Baylor University, Waco, TX 76798-7328, USA\\
%${}^{d}$ EUCOS-CASPER, Physics Department, Baylor University, Waco, TX 76798-7316, USA
%}

\date{\today}

\begin{abstract}

%Starobinsky and $\alpha$-attractor inflation represent two of most favored inflation models with Planck data. 
Recently, the nonperturbative quantization scheme of loop quantum gravity has been extended to the Brans-Dicke theory and the corresponding loop quantum Brans-Dicke cosmology has been derived, which provides an essential platform to explore inflationary models in this framework. In this paper, we consider two inflation models, the Starobinsky and $\alpha$-attractor inflation whose cosmological predictions are in excellent agreement with Planck data, and study systematically their pre-inflationary dynamics as well as  the slow-roll inflation. We show that for both models, the background evolution of a flat Friedmann-Lema\^{i}tre-Robertson-Walker universe in general can be divided into three different phases: the pre-inflationary quantum phase, quantum-to-classical transition, and the slow-roll inflation. The pre-inflationary dynamics are dominated by the quantum geometry effects of loop quantum Brans-Dicke cosmology and the corresponding Universe could be either initially expanding or contracting, depending on the initial velocity of inflaton field. It is shown that the detailed evolution of pre-inflationary quantum phase also depend on specific inflation models. After the pre-inflationary quantum phase, the universe gradually evolves into the slow-roll inflation with some of initial conditions for Starobinsky and $\alpha$-attractor potentials. In addition, to be consistent with observational data, we also find the restricted parameter space of initial conditions that could produce at least $60$ $e$-folds during the slow-roll inflation.

\end{abstract}
\maketitle

%%%%%%%%%%%%%%%%%%%%%%%%%%%%%%%
%%%%%%%%%%%%%%%%%%%%%%%%%%%%%%%
\section{Introduction}
\renewcommand{\theequation}{1.\arabic{equation}}\setcounter{equation}{0}
%%%%%%%%%%%%%%%%%%%%%%%%%%%%%%%
%%%%%%%%%%%%%%%%%%%%%%%%%%%%%%%

The paradigm of cosmic inflation provides perhaps the most compelling picture of the universe at the early stages of its history.  It has achieved remarkable successes not only in solving several problems of the standard big bang cosmology, but most importantly in predicting the primordial perturbations spectra whose evolutions explain both the formation of the large scale structure of the universe and the tiny anisotropies in the cosmic microwave background (CMB) \cite{guth_inflationary_1981, sato_firstorder_1981, starobinsky_new_1980}. All these predictions are all matched to cosmological observational data with high precisions \cite{komatsu_sevenyear_2011, planckcollaboration_planck_2018, planck_collaboration_planck_2014-1, planck_collaboration_planck_2015-4}. In general, there are a lot of approaches to realize the inflation that originate from very different background physics. A common trait of many inflationary models is that they involve scalar degrees of freedom with a self-interacting potential that gives rise to a slow-roll phase during which the energy density of the matter field remains nearly constant and the spacetime behaves like a quasi-de Sitter spacetime. Recently, cosmological and astrophysical data show that for single field inflationary models, predictions of the Starobinsky and $\alpha$-attractor inflation with a small value of $\alpha$ are favored over others \cite{planckcollaboration_planck_2018}. The Starobinsky inflation is based on the account of $R^2$-term as the correction in the Einstein equations \cite{starobinsky_new_1980}, which emerges in the Planck epoch and plays a fundamental role in the high curvature limit, when the early-time acceleration takes place. The $\alpha$-attractor inflation can be in general considered as extensions of the Starobinsky inflation from $\alpha=1$ to $\alpha \neq 1$ \cite{kallosh_superconformal_2013, ferrara_minimal_2013}. Such theories are conformally equivalent to a scalar-tensor theory in the Einstein frame, where the inflaton drives the expansion in a quasi-de Sitter space-time and slowly moves to the end of inflation \cite{tsujikawa_planck_2013, felice_theories_2010}. As expected, their perfect agreement with Planck data has renewed interest in these models.

However, inflationary theory itself is very sensitive to physics at Planck scales, due to the fact that the energy scale of inflation may not be far from that of quantum gravity \cite{martin_transplanckian_2001, zhu_uniform}. Because of this, the underlying quantum field theory and classical general relativity (GR) on classical spacetime becomes unreliable for a large class of inflationary models. This is also known as the ``trans-Planckian issues" of the inflationary theory and its implications on primordial perturbation spectra have also been studied in some concrete quantum theories of gravity, for example see the discussions in Horava-Lifshitz gravity \cite{zhu_hovara}. In addition, insisting on the use of classical GR to describe the inflationary process will inevitably lead to an initial singularity \cite{borde_eternal_1994, borde_inflationary_2003}.  All these issues are closely related to the regime where classical GR is known to break down, and one expects a quantum theory of gravity will provide a completed description of inflation as well as its pre-inflationary dynamics.

To address these issues, Loop quantum cosmology (LQC) provides a natural framework, in which the standard inflationary scenarios can be extended from the onset of the slow-roll inflation back to the Planck era in a self-consistent way. In such a picture, the big bang singularity is replaced by a finite nonzero universe, the quantum bounce, which eventually evolves to the desired slow-roll inflation with very high possibilities \cite{singh_nonsingular_2006, zhang_inflationary_2007, chen_loop_2015, zhu_preinflationary_2017a, agullo_quantum_2012}. Such remarkable features of the quantum bounce have attracted a great deal of attentions lately, in which the universe dominated by a scalar field for different scalar field potentials have been widely explored \cite{ye_loop_2018, zhu_preinflationary_2017a, chen_loop_2015, shahalam_preinflationary_2018, li_qualitative_2018, li_cosmological_2018, shahalam_preinflationary_2017, sharma_preinflationary_2018, agullo_preinflationary_2013, agullo_detailed_2015}. In addition, the primordial perturbations spectra with loop quantum corrections and effects of quantum bounce and their observational constraints have also been discussed extensively (see \cite{zhu_primordial_2018a, zhu_preinflationary_2017a, zhu_universal_2017b, agullo_preinflationary_2013, agullo_loop_2015, agullo_detailed_2015, ashtekar_quantum_2017, zhu_uniform_lqc} and references therein). 

Since the Starobinsky and $\alpha$-attractor inflation models are favored by Planck data, an essential question arising in the framework of LQC is to check if the slow-roll inflation for the  Starobinsky and $\alpha$-attractor models can still occur after the quantum bounce at the Planck era. In fact, both the dynamics of background and cosmological perturbations of Starobinsky inflation model have been already studied in the framework of LQC of GR \cite{bonga_phenomenological_2016, zhu_preinflationary_2017a}. One of main conclusions of these studies is that following the quantum bounce in LQC of GR, a desired slow-roll inflation phase is almost inevitable and the imprints of quantum bounce on primordial perturbation spectra and non-Gaussianities can be well within observational constraints \cite{zhu_universal_2017b, zhu_preinflationary_2017a, zhu_primordial_2018a, wu_nonadiabatic_2018}.

However, as mentioned in \cite{zhu_preinflationary_2017a}, most of the above mentioned studies about Starobinsky inflation are limited to the effective dynamics obtained from the loop quantization in the Einstein frame. In this frame, the original theory of Starobinsky inflation and its extensions ($\alpha$-attractors) in the Jordan frame have been transformed into the Einstein frame by using a conformal transformation, so that the slow-roll inflation can be driven by a scalar field with the specific potentials in the framework of GR. Classically this is correct because the descriptions of the slow-roll inflation in both frames are equivalent. However,  whether this is also true or not in LQC is still an open question. According to \cite{artymowski_comparison_2013}, in general the Einstein and Jordan frames are non longer equivalent at the quantum level. Thus, it is interesting to explore the inflation directly with the effective dynamics obtained from the loop quantization directly in the Jordan frame, based on the quantization proposed in \cite{zhang_loop_2012, zhang_loop_2013a,zhang_extension_2011, zhang_loop_2011,zhang_nonperturbative_2011}. 

In general, both the theories of Starobinsky and $\alpha$-attractor inflation can be casted into the form of specific types of Brans-Dicke (BD) theory in the Jordan frame \cite{tsujikawa_planck_2013, felice_theories_2010}. Recently,  the nonperturbative quantization scheme of LQG has been successfully extended to the BD theory \cite{zhang_loop_2012, zhang_loop_2013a}. The corresponding effective equations of cosmological model for loop quantum BD cosmology have been derived \cite{zhang_loop_2013a} based on the loop quantization procedure in the Jordan frame. These effective equations thus provide a very essential platform to study the Starobinbsky and $\alpha$-attractor inflation in the framework of loop quantum BD cosmology. 

To this purpose, in this paper we study the Starobinsky and $\alpha$-attractor inflation as well as their pre-inflationary dynamics in the framework of loop quantum BD cosmology. Because the quantization in different frames may give different results, we expect loop quantum BD cosmology may provide some distinguishing description about the background evolutions of Starobinsky and $\alpha$-attractor inflation from those in LQC of GR in the Einstein frame. Since most of studies and results about Starobinsky and $\alpha$-attractor inflation are considered in the Einstein frame, here in order to compare our results with theirs, in this paper we shall focus on quantities of background evolution in loop quantum BD cosmology (in Jordan frame) by writing them in terms of those in LQC of GR (in Einstein frame). With this strategy, we show that the evolution of the background in general can be divided into three phases: {\em the pre-inflationary quantum phase, quantum-to-classical transition, and the slow-roll inflation}. For pre-inflationary quantum phase, we shall observe that the Universe starts at a finite non-zero Universe which could be either contracting or expanding depending on the initial velocity of scalar field $\chi$. For slow-roll inflation, we also show the parameter space that could leads to at least $60$ $e$-folds during the slow-roll inflation.

This paper is origanized as follows. In Sec. II, we give an introduction of the classical dynamics of slow-roll inflation in BD theory and in particular focus on the Starobinsky and $\alpha$-attarctor inflation as well as their observational constraints. In Sec. III, we present the effective equations about the background evolution in loop quantum Brans-Dicke cosmology and then transform them in terms of quantities in the Einstein frame. Based on these effective equations, in Sec. IV we turn to study the background evolution by using numerical calculations in details for both Starobinsky and $\alpha$-attractor inflation. Our main conclusions and discussions are presented in Sec. V.

\section{Classical dynamics of Slow-roll inflation in Brans-Dicke theory}
\renewcommand{\theequation}{2.\arabic{equation}}\setcounter{equation}{0}

\subsection{Brans-Dicke theory in Jordan frame}

The action of four dimensional BD theory is given by \cite{brans_mach_1961}
\bqn \lb{BD_action}
S_J= \int d^4 x \sqrt{-g} \Big[\frac{M_{\rm Pl}}{2} \phi R + \frac{M_{\rm Pl}}{\phi} \omega_{\rm BD} X -  V(\phi)\Big],\nb\\
\eqn
where $g$ is the determinant of the spacetime metric $g_{\mu \nu}$, $R$ is the four dimensional Ricci scalar, $\phi$ is the BD scalar field, $\omega_{\rm BD}$ is the BD parameter which is a dimensionless constant, $X \equiv - \frac{1}{2} g^{\mu\nu}(\partial_\mu \phi)( \partial_\nu \phi)$, and $V(\phi)$ represents the potential of the scalar field $\phi$. Note that the gravitational constant $8 \pi G = M_{\rm Pl}^{-2} = 8\pi m_{\rm Pl}^{-2}$ with $M_{\rm Pl}$ and $m_{\rm Pl}$ being the reduced Planck and  Planck mass respectively. We note that in contrary to the original BD theory, we introduced the field potential $V(\phi)$. 

The $f(R)$ theory of gravity with the action
\bqn
S = \int d^4 x \sqrt{-g} \frac{M_{\rm Pl}^2}{2} f(R)
\eqn
 is related to the BD theory by the following correspondence
\bqn
\frac{\phi}{M_{\rm Pl}}= \frac{d f}{dR},\;\; V(\phi)= \frac{M_{\rm Pl}^2}{2} \left(R \frac{df}{dR}-f\right), \;\; \omega_{\rm BD}=0.\nb\\
\eqn
Then the Starobinsky inflation with action \cite{starobinsky_new_1980}
\bqn
S_{R^2} = \int d^4 x \sqrt{- g} \frac{M_{\rm Pl}^2}{2} \left(R + \frac{R^2}{6M^2}\right)
\eqn
is a special case of the Brans-Dicke theory with
\bqn
\frac{\phi}{M_{\rm Pl}} = 1+ \frac{R}{3M^2}, \;\;\\
V(\phi) = \frac{3M^2}{4}(\phi-M_{\rm Pl})^2,\\
\omega_{\rm BD}=0.
\eqn

In the Jordan frame, we consider the spatially flat Friedmann-Lema\^{i}tre-Robertson-Walker (FLRW) universe, for which the metric takes the form
\bqn
ds^2 = - dt^2 + a^2(t) dx^i dx_i,
\eqn
where $a(t)$ is the scale factor of the Universe and $t$ is the cosmic time in the Jordan frame. The dynamics of the universe can be derived from the action (\ref{BD_action}), which is governed by the modified Friedmann and Klein-Gordon equations, i.e., 
\bqn
\lb{friedmann}
3M_{\rm Pl}^2 \left(H + \frac{\dot \phi}{2 \phi}\right)^2=
% \frac{\beta}{2}\frac{M_{\rm Pl}^2}{\phi^2} \frac{\dot \phi^2}{2}+ \frac{M_{\rm Pl}}{\phi}V(\phi) =
  \frac{M_{\rm Pl}^2 \rho_\phi}{\phi^2},\\
\lb{kg}
\ddot \phi + 3 H \dot \phi +\frac{2}{\beta M_{\rm Pl}}\Big(\phi V_{\phi}(\phi) - 2 V(\phi)\Big)=0.
\eqn
where $H\equiv \dot a /a$ denotes the Hubble parameter, $\beta \equiv 2 \omega_{\rm BD} +3$, and $\rho_{\phi} \equiv \frac{\beta}{4} \dot \phi^2 + \phi V(\phi)/M_{\rm Pl}$ is the effective energy density of the BD scalar field.
%, which relate to those in the Einstein frame as
%\bqn
%a(t) = \sqrt{\frac{M_{\rm Pl}}{\phi}} \hat a(\hat t), \;\; dt =  \sqrt{\frac{M_{\rm Pl}}{\phi}} d \hat t.
%%(t, \; x^i) = \sqrt{\frac{\phi}{M_{\rm Pl}}} (\hat t, \; \hat x^i )
%\eqn

\subsection{Starobinsky inflation and $\alpha$-attractors in Einstein frame}

Under the conformal transformation 
\bqn\lb{conformal}
\hat g_{\mu\nu} = \frac{\phi}{M_{\rm Pl}} g_{\mu\nu},
\eqn
the action (\ref{BD_action}) can be recasted to the one with a minimally coupled scalar field $\chi$ in the Einstein frame. The transformed action is given by
\bqn\lb{SE}
S_{\rm E} = \int d^4 x \sqrt{-\hat g} \left[\frac{M_{\rm Pl}^2}{2}\hat R - \frac{1}{2}\hat g^{\mu\nu} \partial_\mu \chi \partial_\nu \chi -U(\chi)\right],\nb\\
\eqn
where a hat represents the quantities in the Einstein frame, and one can identify 
\bqn
U(\chi) &=& e^{-2 \sqrt{\frac{2}{\beta}} \frac{\chi}{M_{\rm pl}}} V(\phi),\;\;\\
\frac{\phi}{M_{\rm Pl}} &=& e^{\sqrt{\frac{2}{\beta}} \frac{\chi}{M_{\rm Pl}}}.
\eqn
For Starobinsky inflation, it corresponds to the scalar field $\chi$ and the corresponding Starobinsky potential $U(\chi)$ as
\bqn
\frac{\chi}{M_{\rm Pl}} &=&\sqrt{\frac{3}{2}}\ln\left(\frac{\phi}{M_{\rm Pl}}\right)= \sqrt{\frac{3}{2}}\ln\left(1+\frac{R}{3M^2}\right),\;\\
U(\chi) &=& \frac{3}{4}M^2 M_{\rm Pl}^2 \left(1-e^{-\sqrt{\frac{2}{\beta}} \frac{\chi}{M_{\rm Pl}}}\right)^2, \lb{starobinsky}\\
\beta&=&3.
\eqn
The Starobinsky potential $U(\chi)$ with $\beta=3$ can be extended to $\beta\neq 3$, which corresponds to a subclass of the E-type $\alpha$-attractor inflation models \cite{kallosh_superconformal_2013, ferrara_minimal_2013, galante_unity_2015}. Normally, the action of the $\alpha$-attractor inflation models of a real scalar field $\varphi$ can be written in the Einstein frame in the non-canonical form \cite{kallosh_superconformal_2013, ferrara_minimal_2013, galante_unity_2015}
\bqn
S&=&\int d^4 x \sqrt{-\hat g}\Bigg(\frac{M_{\rm Pl}^2}{2} \hat R + \frac{1}{2}\frac{\alpha}{\left(1- \frac{\varphi^2}{6 M_{\rm Pl}^2}\right)^ 2} \hat g^{\mu\nu} (\partial_\mu \varphi) \partial_\nu \varphi  \nb\\
&&~~~~~~~~~~~ + f^2\left(\frac{\varphi}{\sqrt{6}M_{\rm Pl}}\right)\Bigg),
\eqn
where $f(\varphi/(\sqrt{6}M_{\rm Pl}))$ denotes an arbitrary function. This action can also be described by the canonical action (\ref{SE}) by redefining a canonical scalar field $\chi$ as
\bqn
\frac{\varphi}{\sqrt{6} M_{\rm Pl}} = {\rm tanh}\left(\frac{\chi}{\sqrt{6 \alpha} M_{\rm Pl}}\right).
\eqn
Different choices of the arbitrary function $f$ correspond to the different types of inflation attractors. The E-type $\alpha$-attractor which we considered in this paper corresponds to the function $f(\varphi/(\sqrt{6}M_{\rm Pl}))$ with a special choice such that
\bqn
f^2\left[{\rm tanh}\left(\frac{\chi}{\sqrt{6 \alpha} M_{\rm Pl}}\right)\right] = U(\chi),
\eqn
where $U(\chi)$ is given by (\ref{starobinsky}) with $\alpha=\beta/6$. 
In the Jordan frame, this case corresponds to the BD theory with $\omega_{\rm BD} \neq 0$. The above analysis shows that the Starobinsky and the E-type $\alpha$-attractor inflation models can be conformally equivalent to the classical BD cosmology with $\omega_{\rm BD}=0$ and $\omega_{\rm BD} \neq 0$ respectively. 

Now let us turn to consider the dynamics of a spatially flat FLRW universe in the Einstein frame, 
\bqn
ds^2 = - d \hat t^2 +\hat a^2(\hat t) d x^i d x_i
\eqn
with $\hat a(\hat t)$ being the scale factor of the universe and $\hat t$ is the cosmic time in Einstein frame. The dynamics of the background cosmology can be derived from the action (\ref{SE}), which is governed by
\bqn
\hat H^2 = \frac{1}{3M_{\rm Pl}^2} \hat \rho_\chi, \lb{C_E_F}\\
\frac{d^2\chi}{d\hat t^2}+3 \hat H \frac{d\chi}{d \hat t} +U_\chi=0,\lb{C_E_KG}
\eqn
where
\bqn
\hat \rho_\chi \equiv \frac{1}{2}\dot \chi^2 +U(\chi),\\
U_\chi \equiv \frac{dU(\chi)}{d\chi}.
\eqn

To study the slow-roll inflation, it is convenient to introduce the two slow-roll conditions
\bqn
\frac{1}{2}\dot \chi^2 \ll U(\chi),\\
\left|\frac{d^2 \chi}{d\hat t^2}\right| \ll \left|3 \hat H \frac{d\chi}{d \hat t}\right|, \left|U_\chi\right|.
\eqn
Then the Friedmann and Klein-Gordon equations become
\bqn
3M_{\rm Pl}^2 \hat H^2 \simeq U(\chi),\\
3 \hat H \frac{d\chi}{d \hat t} \simeq -U_\chi.
\eqn
With these two approximate equations, the e-folds during the slow-roll inflation reads
\bqn
N_{\rm inf} &=& \int_{\hat a_{\rm i}}^{\hat a_{\rm end}} \frac{d\hat a }{\hat a} =  \int_{\hat t_i}^{\hat t_{\rm end}} \hat H d\hat t = - \int_{\chi_{\rm end}}^{\chi_{i}} \frac{\hat H}{d\chi/d\hat t} d\chi \nb\\
&\simeq& \frac{1}{M_{\rm Pl}^2}\int_{\chi_{\rm end}}^{\chi_{i}} \frac{ U(\chi)}{U_\chi(\chi)} d\chi.
\eqn
To describe the evolution of the slow-roll background, we introduce the Hubble slow-roll parameter $\epsilon_H$ and potential slow-roll parameter $\epsilon_U$ as,
\bqn
\epsilon_H \equiv - \frac{\dot H}{H^2}, \;\; \epsilon_U \equiv M_{\rm Pl}^2 \frac{ U_\chi^2}{2U^2}.
\eqn
During the slow-roll inflation, up to the leading-order in the slow-roll approximation, we have
\bqn
\epsilon_H \simeq \epsilon_U.
\eqn

For the Starobinksy ($\beta=3$) and $\alpha$-attractor ($\beta\neq 3$) inflation, the potential of the scalar field is given by (\ref{starobinsky}), from which one obtains
\bqn
\epsilon_U \simeq = \frac{4}{\beta} \left(1- e^{\sqrt{\frac{2}{\beta}} \frac{\chi}{M_{\rm Pl}}}\right)^{-2},
\eqn 
and
\bqn
N_{\rm inf} \simeq \frac{\beta}{4}\left(e^{\sqrt{\frac{2}{\beta}} \frac{\chi_{i}}{M_{\rm Pl}}}-e^{\sqrt{\frac{2}{\beta}} \frac{\chi_{\rm end}}{M_{\rm Pl}}}\right) +\frac{1}{2}\sqrt{\frac{\beta}{2}}\frac{\chi_{\rm end}-\chi_i}{M_{\rm Pl}}.\nb\\
\eqn
Then at the end of the slow-roll inflation, $\epsilon_U \simeq 1$ yields
\bqn
\frac{\chi_{\rm end}}{M_{\rm Pl}} = \sqrt{\frac{\beta}{2}} \ln\left(1+\sqrt{\frac{4}{\beta}}\right).
% \simeq 0.94.
\eqn
To produce at least $60$ e-folds during the slow-roll inflation (i.e. $N_{\rm inf} = 60$), we have
\bqn
\frac{\chi_{i}}{M_{\rm Pl}} = - \sqrt{\frac{\beta}{2}} \left(c+W_{-1}(-e^{-c})\right),
% 5.45.
\eqn 
with $c$ being given by
\bqn
c \equiv \frac{4}{\beta} N_{\rm inf} + e^{\sqrt{\frac{2}{\beta}} \frac{x_{\rm end}}{M_{\rm Pl}}} - \sqrt{\frac{2}{\beta}} \frac{x_{\rm end}}{M_{\rm Pl}},
\eqn
and $W_{-1} (x)$ denoting the Lambert $W$ function. 

The inflationary observables of Starobinsky and $\alpha$-attractor inflation are given by
\bqn
n_s \simeq 1 - \frac{2}{N_{\rm inf}},\;\; r \simeq \frac{4 \beta}{N_{\rm inf}^2},
\eqn
where $n_s$ is the spectral index of the inflationary scalar power spectrum and $r$ is the ratio between the tensor and scalar spectra. Then the Planck 2018 data together with the BICEP2/Keck Array 2014 data set tight constraint on $\beta$ \cite{planckcollaboration_planck_2018},
\bqn\lb{beta_constraint}
\beta \lesssim 94.87 \;\;\; {\rm at \;\; 95\% \;\; C.L.}.
\eqn

\subsection{Equivalence between Einstein and Jordan frames}

At the classical level, both the Einstein and Jordon frames can be equivalently transformed to each other. According to the conformal transformation, one can establish relationships between the quantities in both frames, 
\bqn
\hat a (\hat t) &=& \sqrt{\frac{\phi}{M_{\rm Pl}}} a( t), \;\; d \hat t =  \sqrt{\frac{\phi}{M_{\rm Pl}}} d t,\\
%(t, \; x^i) = \sqrt{\frac{\phi}{M_{\rm Pl}}} (\hat t, \; \hat x^i )
\hat H &=& \sqrt{\frac{M_{\rm Pl}}{\phi}} \left(H+ \frac{\dot \phi}{2 \phi}\right), \lb{H}\\
\frac{d\chi}{d\hat t} &=& \sqrt{\frac{\beta}{2}} \left ( \frac{M_{\rm Pl}}{\phi} \right)^\frac{3}{2} \dot \phi, \lb{chi_dot}\\
\frac{d^2\chi}{d\hat t^2} &=& \sqrt{\frac{\beta}{2}} M_{\rm Pl}^2 \left( \frac{\ddot \phi}{\phi^2} - \frac{3 \dot \phi^2}{2 \phi^3}\right), \lb{chi_ddot}\\
\hat \rho_\chi &=& \left ( \frac{M_{\rm Pl}}{\phi} \right)^3 \left ( \frac{\beta}{4} \dot \phi^2 + \frac{\phi}{M_{\rm Pl}}V(\phi)\right).  \lb{rho_chi}
\eqn
With the above identifications, one can easily verify that the dynamics of the BD cosmology in both Einstein and Jordan frame are equivalent to each other.

\section{Effective equations in loop quantum Brans-Dicke cosmology}
\renewcommand{\theequation}{3.\arabic{equation}}\setcounter{equation}{0}

Loop quantum gravity (LQG) provides a background independent way to quantize GR, which has been widely investigated in the past 30 years \cite{ashtekar_background_2004, rovelli_quantum_2007, han_fundamental_2007}. In the framework of LQG, it is remarkable that GR can be non-perturbatively quantized by the loop quantization procedure. Recently, this promising loop quantization has been extended to theories of modified gravity, for examples,  BD theory \cite{zhang_loop_2012, zhang_loop_2013a}, metric $f(R)$ theories \cite{zhang_extension_2011, zhang_loop_2011}, and scalar-tensor theories \cite{zhang_nonperturbative_2011}.  Applying the physical ideas and mathematical tools underlying LQG, LQC is proposed, which represents a symmetry-reduced model of LQG by quantizing the FLRW background spacetime with the loop quantization procedures (for review of LQC see \cite{ashtekar_loop_2011, bojowald_loop_2005} and references therein).

For cosmology of BD theory, the background dynamics can be described in two ways, direct investigation in the Jordan frame and using certain conformal transformation to the Einstein frame. It has been shown that the results derived from the LQC quantization in different frames are not equivalent to each other \cite{artymowski_comparison_2013}. Thus it is interesting to explore the effective dynamics of cosmology with quantization in different frames. With the effective equations derived from the LQC quantization in the Einstein frame, the effective cosmological dynamics have been studied extensively in both frames (see \cite{artymowski_comparison_2013, bojowald_loop_2006, bojowald_singularities_2006, bojowald_loop_2006, artymowski_inflation_2013, bojowald_singularities_2006} for examples). It is worth noting that in \cite{artymowski_inflation_2013, bojowald_singularities_2006} the dynamics of the slow-roll inflation from non-minimal coupled scalar field have been discussed by conformally transforming the LQC quantization results in the Einstein frame to Jordan frame. In this paper, in contrast to these works, we consider the cosmological evolution of BD cosmology by directly using the effective dynamics from LQC quantization in the Jordan frame.

The effective dynamics of loop quantum BD cosmology is derived in the Jordan frame in Refs.~\cite{zhang_loop_2013, artymowski_comparison_2013}, in which the effective equations of both the Friedmann and Klein-Gordon equations for the background cosmology with the BD scalar field $\phi$ are given by
\bqn
&&\left(H + \frac{\dot \phi}{2\phi}\right)^2 \nb\\
&&~~~~ = \left(\frac{1}{\phi}\sqrt{\frac{\rho_\phi}{3}} \sqrt{1-\frac{\rho_\phi}{\rho_{\rm c}}} +\frac{\dot \phi}{2 \phi} \left(1-\sqrt{1-\frac{\rho_\phi}{\rho_{\rm c}}}\right)\right)^2, \nb \lb{fri_bd}\\
&&\ddot \phi+ 3 H \dot \phi  +\frac{2}{\beta M_{\rm Pl}} \phi V_\phi  \nb\\
&&~~~~~~ +\frac{2}{\beta M_{\rm Pl}} V(\phi) \left(1- 3 \sqrt{1-\frac{\rho_\phi}{\rho_{\rm c}}}\right)=0. \lb{kg_bd}
\eqn
It is obvious to see that the effective energy density of the BD field $\rho_{\phi}$ now has a maximum value $\rho_{\rm c}$. When $\rho_{\phi}$ approaches this maximum value, the Hubble parameter $H$ in the Jordan frame approaches zero, which implies a quantum bounce occurs at this point in the Jordan frame. As a result, the past singularity arising in the classical BD universe in the Jordan frame is cured by the quantum bounce. When $\rho_{\phi} \ll \rho_{\rm c}$, the above equations reduce to the classical version in the classical BD theory. 

Note that our purpose is to study Starobinsky and $\alpha$-attractor inflation in quantum BD cosmology. Since most of studies and results about this two inflation models are considered in the Einstein frame, here in order to compare our results with theirs, we shall focus on quantities in terms of variables such as $\hat H$ and $\chi$ in the Einstein frame. By this treatment, we consider the Einstein frame as the physical frame. In the Einstein frame, By using the relations in Eqs.~(\ref{H} - \ref{rho_chi}), one obtains
\bqn
\hat H^2 = \frac{1}{M_{\rm Pl}^2} \left( \sqrt{\frac{\hat \rho_\chi}{3}}\sqrt{1 - r_\chi} + \sqrt{\frac{1}{2 \beta}} \dot \chi \left( 1-\sqrt{1 - r_\chi} \right) \right)^2, \lb{lqc_J_E_F} \nb\\
\eqn
and
\bqn
\frac{d^2 \chi}{d \hat t^2} + 3 \hat H \frac{d\chi}{d \hat t} + U_{\chi} + 3 \sqrt{\frac{2}{\beta}} \frac{U}{M_{\rm Pl}}\left( 1-\sqrt{1 - r_\chi} \right)=0, \nb\\\lb{lqc_J_E_KG} 
\eqn
where 
\bqn
r_\chi \equiv e^{3 \sqrt{\frac{2}{\beta}} \frac{\chi}{M_{\rm Pl}}}\frac{\hat \rho_\chi}{\rho_{\rm c}}.
\eqn
We note that Eqs.~(\ref{lqc_J_E_F}) and (\ref{lqc_J_E_KG}) obtain the classical limit for $r_\chi \ll 1 $, which is described by Eqs.~(\ref{C_E_F}) and (\ref{C_E_KG}). Unlike in LQC of GR that the energy density $\hat \rho_{\chi}$ has a maximum value $\rho_{\rm c}$, now it is restricted to
\bqn
r_{\chi} \leq 1,
\eqn
which leads to
\bqn
e^{3 \sqrt{\frac{2}{\beta}}  \frac{\chi}{M_{\rm Pl}}} \hat \rho_{\chi} < \rho_{\rm c}.
\eqn
It is worth pointing out that the description of the dynamics by using the quantities in the physical Einstein frame is very different from that in the Jordan frame. As mentioned, the quantum bounce in the Jordan frame defined by $H=0$ appears for $\rho_{\phi}=\rho_{\rm c}$, while at this moment we have $r_\chi=\rho_{\phi} / \rho_{\rm c}=1$ in the physical Einstein frame and Eq.(\ref{lqc_J_E_F}) gives $\hat H^2 = \frac{\dot \chi^2}{2 \beta M_{\rm Pl}^2} \neq 0$. The quantum bounce in the physical Einstein frame is defined by $\hat H =0 $ and can be obtained by solving Eqs.~(\ref{lqc_J_E_F}) and (\ref{lqc_J_E_KG}) as we shown later in the next section. This implies the corresponding starting point of Universe is not at the bounce point, but instead by a finite Universe which is either expanding or contracting. This definitely provides another resolution of the initial singularity and its novel properties are still waiting for exploring. 

It should be noted that the above upper bound of the energy density $\rho_{\rm c}$ is in the Jordan frame. By transforming into the Einstein frame, the upper bound becomes $e^{3 \sqrt{\frac{2}{\beta}}  \frac{\chi}{M_{\rm Pl}}} \rho_{\rm c}$. This means the initial condition for the BD field $\phi$ in the Jordan frame and its counterpart scalar field $\chi$ in the Einstein frame can be significantly different.  Normally, one expects that the natural initial condition for inflation when the universe first emerged from the big bang and reached the Planck density is such that all different energy forms (kinetic, gradient, and potential energy) are of the same order. This assumption, as observed in \cite{ijjas_inflationary_2013}, creates a ``unlikeness" problem for inflation models with plateau-like potentials (for example the Starobinsky potential), for which the initial potential energy has to be much smaller than the Planck energy for the occurrence of the slow-roll inflation. Such issue has also been addressed earlier by comparing the initial conditions in both the Jordan and Einstein frames \cite{gorbunov_are_2015}, in which it is shown that when $\hat \rho_{\chi}$ in Einstein frame is at the energy scale $\sim 10^{-12} m_{\rm Pl}^4$ for the occurrence of the slow-roll inflation, $\rho_\phi$ in the Jordan frame could be at Planck scale $\sim m_{\rm Pl}^4$. This fact, as pointed out in \cite{gorbunov_are_2015}, can relieve the ``unlikeness" problem to inflation models with plateau-like potentials that observed in \cite{ijjas_inflationary_2013}. In the context of LQC, this initial condition issue may be different from that of classical inflation in two ways. First, comparing to the classical theory, the dynamics of the universe at the Planck scale is dramatically changed in which a quantum bounce occurs at the Planck era. The Universe starting at the quantum bounce is followed by a super-inflation period before it enters into the classical regime \cite{Ashtekar:2009mm}. Second, the natural initial condition that should be set at the Planck scale with various energy forms being at the same order is in general considered as an assumption and its specific form is expected to be determined by the quantum theory of gravity. As one of candidates of quantum theory of gravity, the loop quantization for the BD cosmology in the Jordan frame directly implies an upper bound on the energy density $\rho_\phi \leq \rho_{\rm c}$. As we mentioned above, the quantum bounce in the Jordan frame occurs when $\rho_\phi = \rho_{\rm c} \sim m_{\rm Pl}^4$ which thus could provide a concrete realization for the natural initial condition for inflation in loop quantum BD cosmology.

On the other hand, by comparing the loop quantization in two different frames, one observes that the quantization directly implies upper bound on the energy density in the same frame that used to implement the quantization procedure. For the quantization in Einstein frame, the energy density is restricted to $\hat \rho_{\chi}\leq \hat \rho_{\rm c}$, while for the quantization in the Jordan frame the upper bound becomes $\rho_\phi \leq \rho_{\rm c}$. The critical energy $\hat \rho_{\rm c}$ and $\rho_{\rm c}$ are in principle different, but both are supposed to be at Planck scale. This fact, as we shall shown in the next section, can directly affect the maximal value of $\chi$ and the initial conditions that can produce at least $60$ e-folds.

\section{Numerical analysis for the Background evolution of Starobinsky and $\alpha$-attractor inflation}
\renewcommand{\theequation}{4.\arabic{equation}}\setcounter{equation}{0}

\begin{table}
\renewcommand\arraystretch{1.5}
	\centering
	\caption{Table for $\beta$, $M$ and $\chi_{\rm max} $. We only show the values which shall be used in the following subsection.}\label{beta_M}
	\begin{tabular}{p{0.11\textwidth}<{\centering}p{0.11\textwidth}<{\centering}p{0.11\textwidth}<{\centering}}
		\hline
\hline
$\beta $  	  &      $M/m_{\rm Pl}$                 &    $\chi_{\rm max}/m_{\rm Pl}$    \\ \hline
1	        &      $1.37\times10^{-6}$           &         1.39             \\
3            &      $2.31\times10^{-6}$           &         2.32             \\
5            &      $2.92\times10^{-6}$           &         2.95            \\
10           &      $3.99\times10^{-6}$           &         4.08            \\
20           &      $5.38\times10^{-6}$           &         5.65            \\
\hline
\hline
	\end{tabular}
\end{table}

In this section, we start to study the evolution of the Universe in the framework of loop quantum BD cosmology by obtaining the solutions of Eqs.~(\ref{lqc_J_E_F}) and (\ref{lqc_J_E_KG}). These equations can be solved numerically by imposing initial conditions for $\hat a(\hat t)$, $\chi (\hat t)$, and $ \dot \chi (\hat t)$ at a specific time. A convenient choice of such a point is at the time $\hat t_0$ when $r_{\chi} =1$, at which we have the two relations
\bqn
\frac{1}{2}\dot \chi_0^2+ U(\chi_0) = \rho_{\rm c} e^{- 3 \sqrt{\frac{2}{\beta}} \frac{\chi_0}{M_{\rm Pl}}},
\eqn
and
\bqn
\hat H_0^2 = \frac{\dot \chi_0^2}{2 \beta M_{\rm Pl}^2}.
\eqn
For the sake of simplicity, we rescale $\hat a(\hat t)$ by setting $\hat a(\hat t_0)=1$ at $\hat t= \hat t_0$. Once the potential $U(\chi)$ is specified, we can obtain $\dot \chi_0$ in terms of $\chi_0$ and $\rho_{\rm c}$ by using the first relation for both $\dot \chi_0 >0$ and $\dot \chi_0 <0$ respectively. In this sense, we only need to specify the value of $\chi_0$ and the sign of $\dot \chi_0$ as initial conditions. In addition, the first relation in the above also restricts $\chi_0$ to the range of $(-\infty, \chi_{\rm max})$, where $\chi_{\rm max}$ can be obtained from
\bqn
e^{ - \sqrt{\frac{2}{\beta}} \frac{\chi_{\rm max}}{M_{\rm Pl}}}=
\frac{M^2 M_{\rm Pl}^2}{4 \rho_{\rm c}} \left( 1- \mathcal{Y}^{\frac{1}{3}}-  \mathcal{Y}^{- \frac{1}{3}} \right)+2  \mathcal{Y}^{- \frac{1}{3}}, \nb\\
\eqn
where
\bqn 
\mathcal{Y} & \equiv & -1+ \frac{12 \rho_{\rm c}}{M^2 M_{\rm Pl}^2} - \frac{24 \rho_{\rm c}^2}{M^4 M_{\rm Pl}^4} \nb\\
&&+8 \left( \frac{ \rho_{\rm c}}{M^2 M_{\rm Pl}^2} \right)^{\frac{3}{2}} \sqrt{\frac{9\rho_{\rm c}}{M^2 M_{\rm Pl}^2}-1}.
\eqn
In Table.~\ref{beta_M}, we present several values of $\chi_{\rm max}$ by several sets of values for $\beta$ and $M$ we used in the numerical calculations. 

From the second relation, we observe that the universe is not at the bounce initially since $\hat H_0 \neq 0$. Depending on the sign of $\dot \chi_0$,  the Universe could be either expanding or contracting at the initial time. Since at this moment there is a quantum bounce in the description in the Jordan frame, i.e., $H=0$, from Eq.~(\ref{H}) we see that the sign of $\hat H_0$ is the same as $\dot \phi_0$, which also has the same sign as $\dot \chi_0$ according to (\ref{chi_dot}). Therefore, at the initial time, the Universe is expanding if $\dot \chi_0 >0$ and contracting when $\dot \chi_0 <0$. Then we have
\bqn
\hat H_0 = \frac{\dot \chi_0}{\sqrt{2 \beta } M_{\rm Pl}}.
\eqn

In order to do the numerical calculation, we also need to specify the values of $\beta$ and $M$. In fact, for each value of $\beta$ within the constraint given by (\ref{beta_constraint}), the values of $M$ can be determined by using the most recent released Planck 2018 data \cite{planckcollaboration_planck_2018}. We present the values of $M$ for $\beta=1, \; 3, \; 5,\; 10,\; 20$ in Table.~{\ref{beta_M}}. 

In the following subsections, we shall study the evolution of the background for potential $U(\chi)$ in (\ref{starobinsky}) for several different values of $\beta$ respectively.

\subsection{Starobinsky potential}

\begin{figure}
\centering
\includegraphics[width=8.5cm]{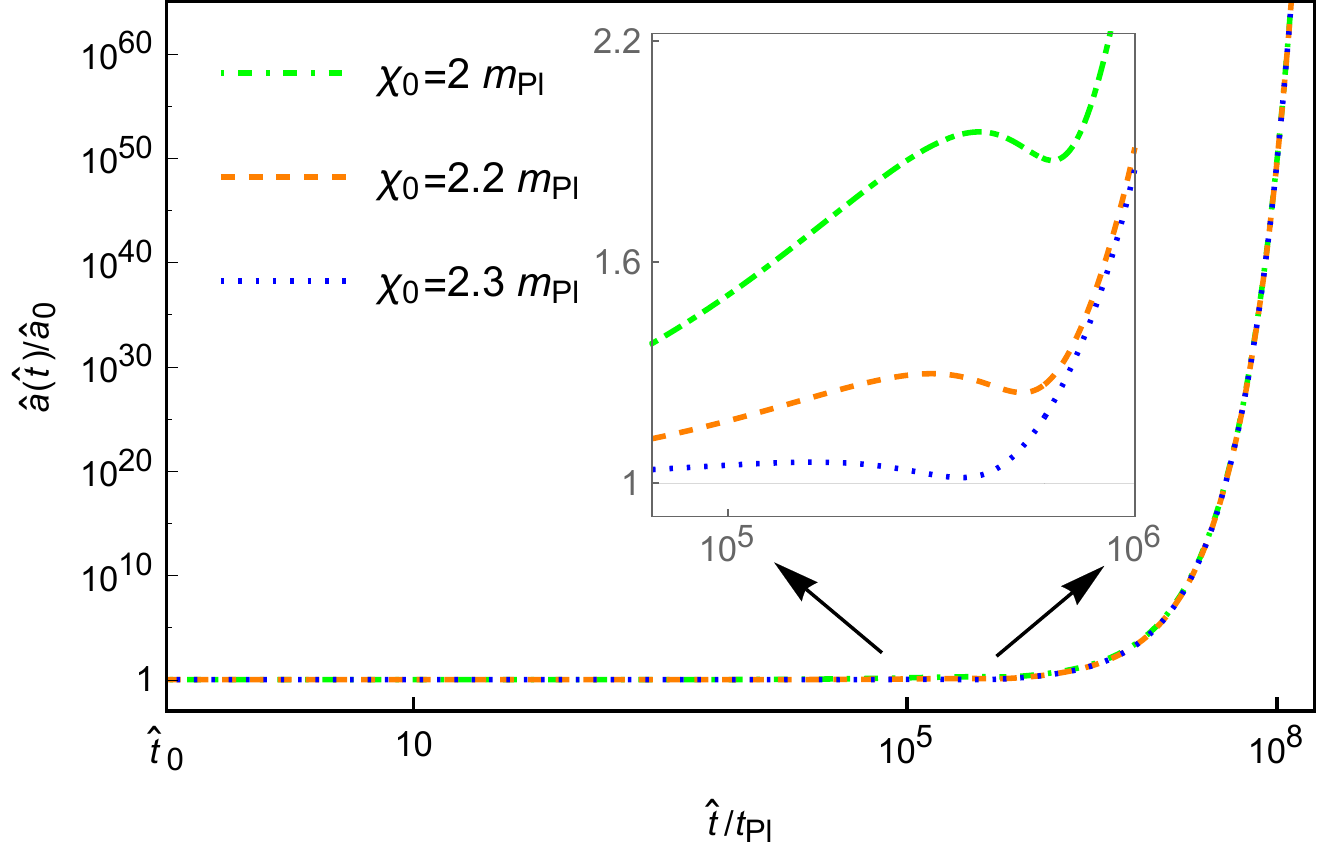}
\includegraphics[width=8.5cm]{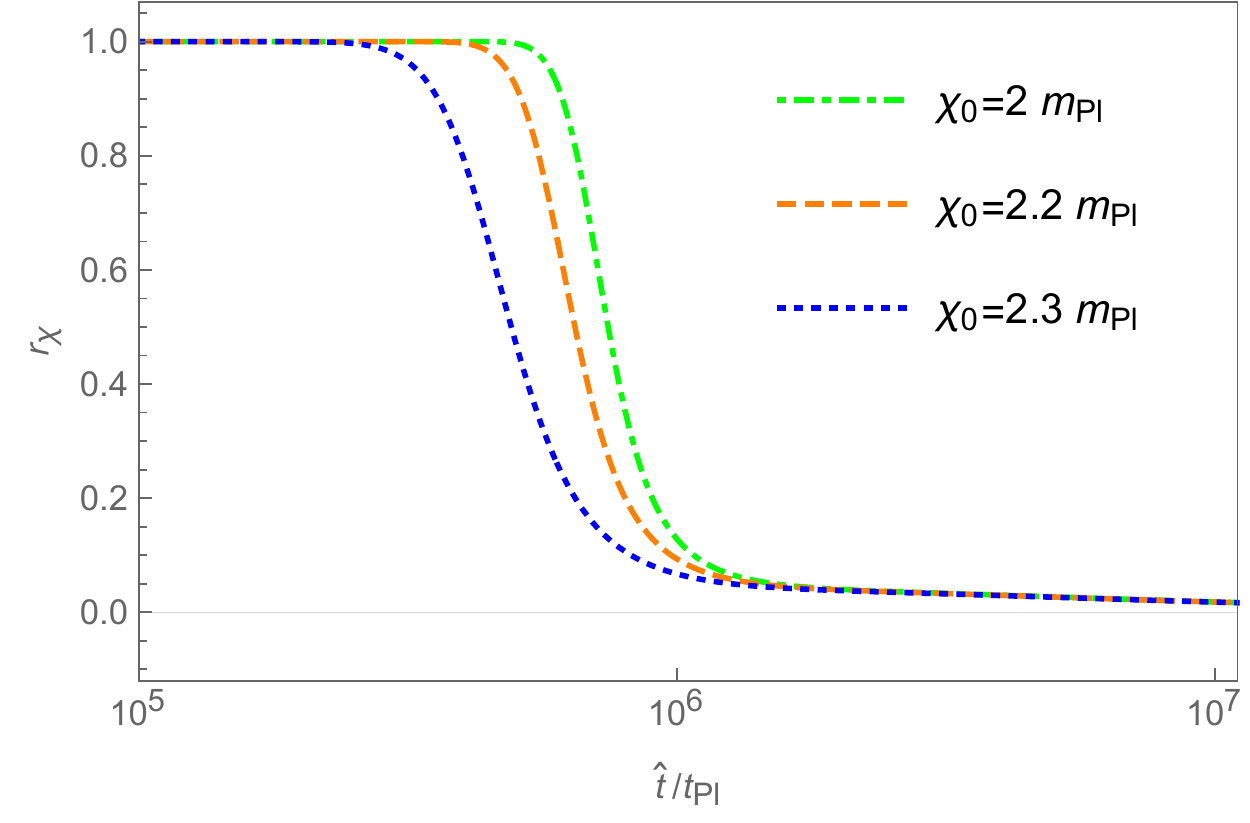}
\includegraphics[width=8.5cm]{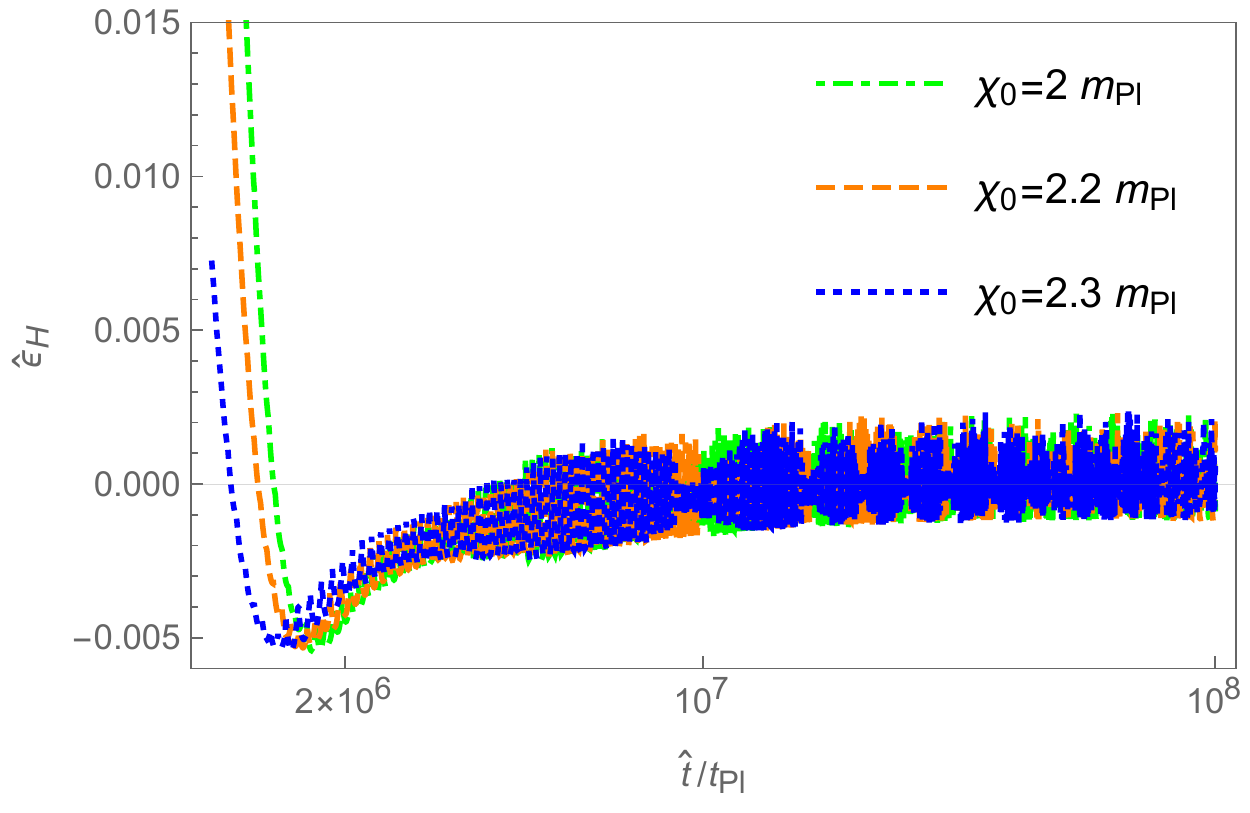}
\caption{Numerical solution for Starobinsky inflation ($\beta=3$) in loop quantum BD cosmology with $\dot \chi_0>0$. Top panel: The evolution of the scale factor $\hat a (\hat t)$ for different initial values of $\chi_0$. Middle panel: $r_\chi$ for the same set of initial conditions. Bottom panel: The evolution of the slow-roll parameter $\hat \epsilon_{H}$.}
\label{dot_chi>0}
\end{figure}

\begin{figure}
\centering
\includegraphics[width=8.5cm]{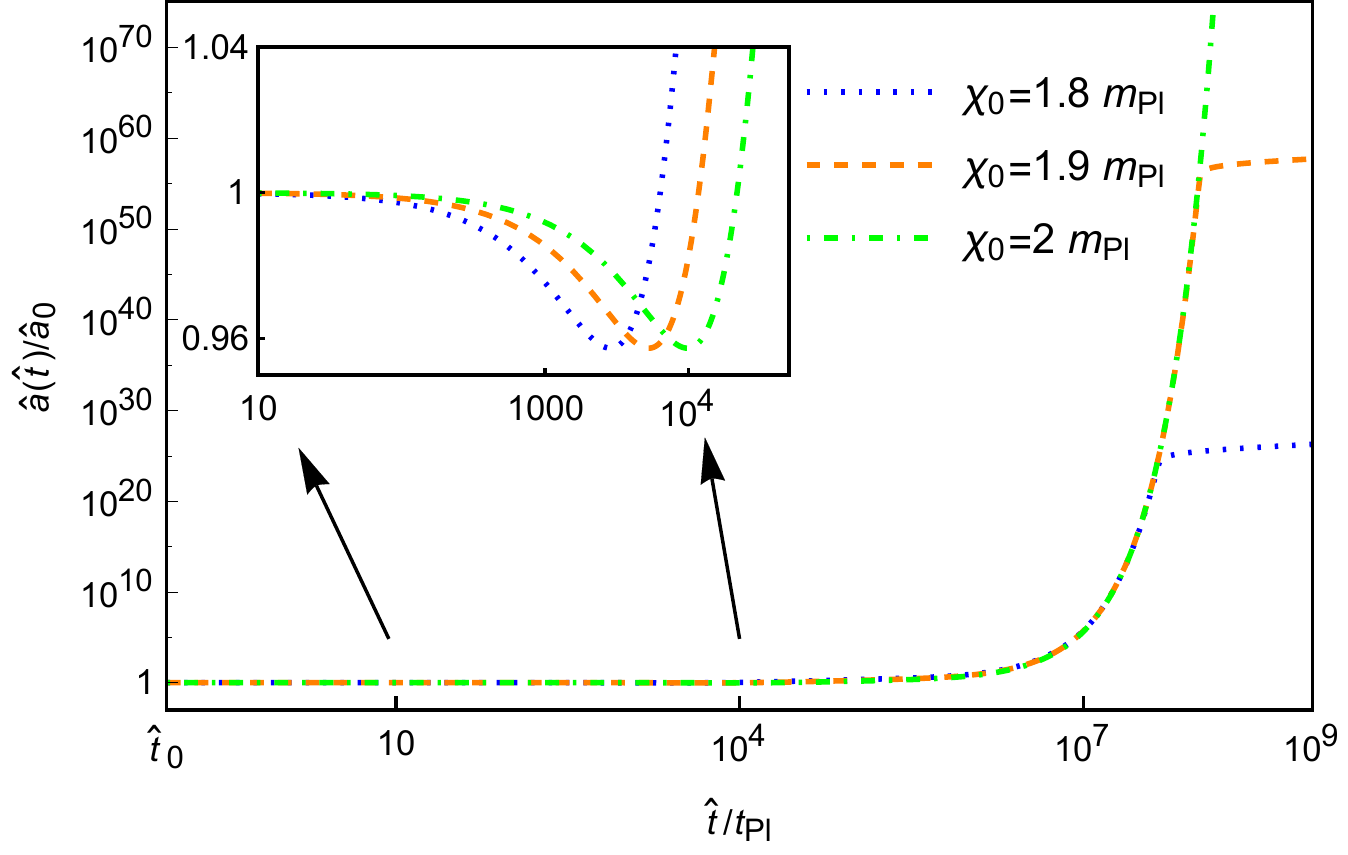}
\includegraphics[width=8.5cm]{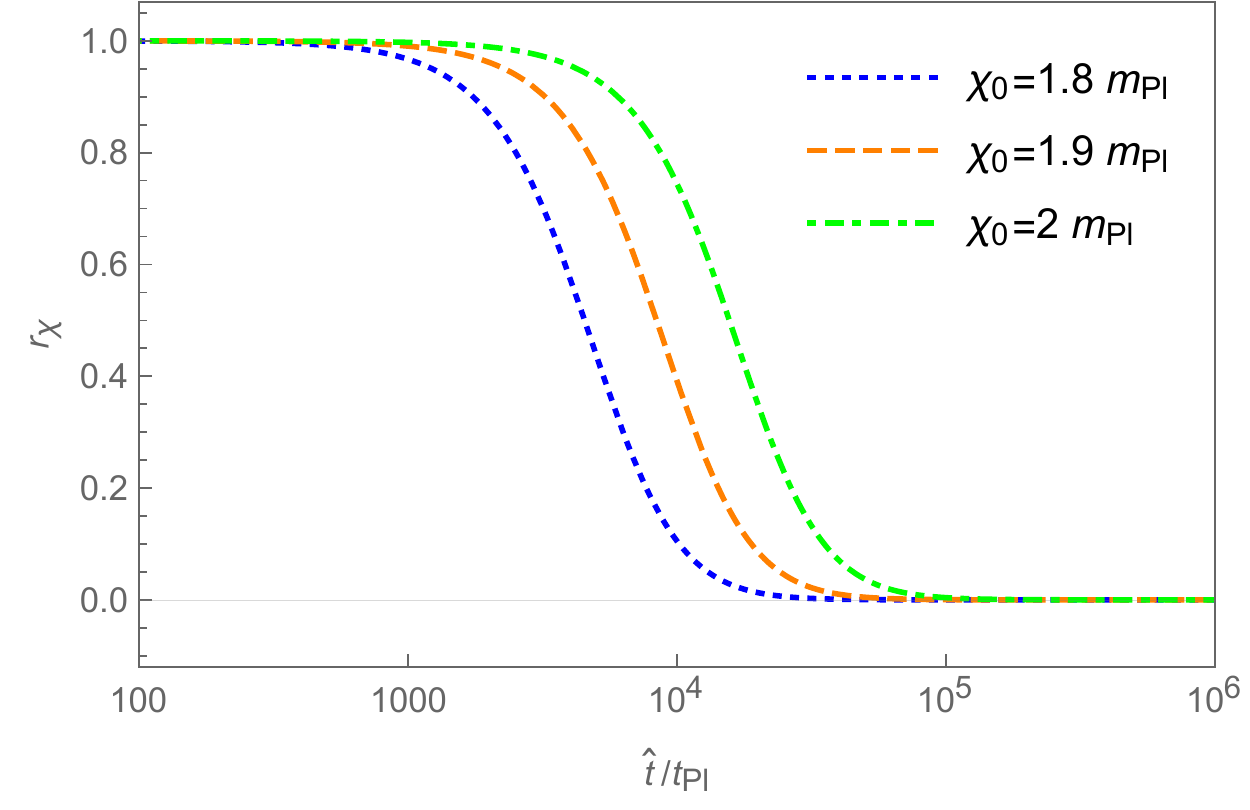}
\includegraphics[width=8.5cm]{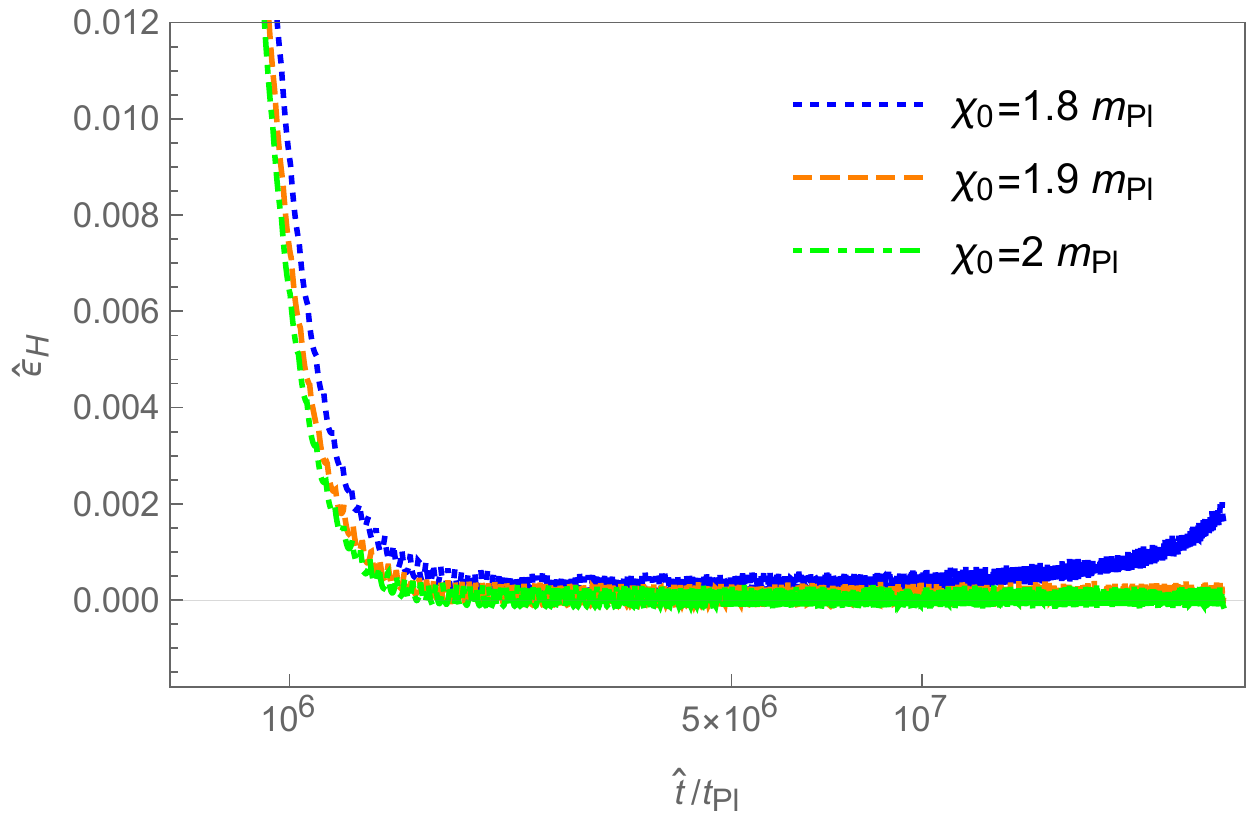}
\caption{Numerical solutions of scale factor $\hat a( \hat t)$  (top panel), $r_{\chi}$  (middle panel), and the slow-roll parameter $\epsilon_{H}$ (bottom panel) for Starobinsky inflation ($\beta=3$) in loop quantum BD cosmology with $\dot \chi_0<0$. }
\label{dot_chi<0}
\end{figure}

 \begin{figure}
  \centering
    \includegraphics[width=.46\textwidth]{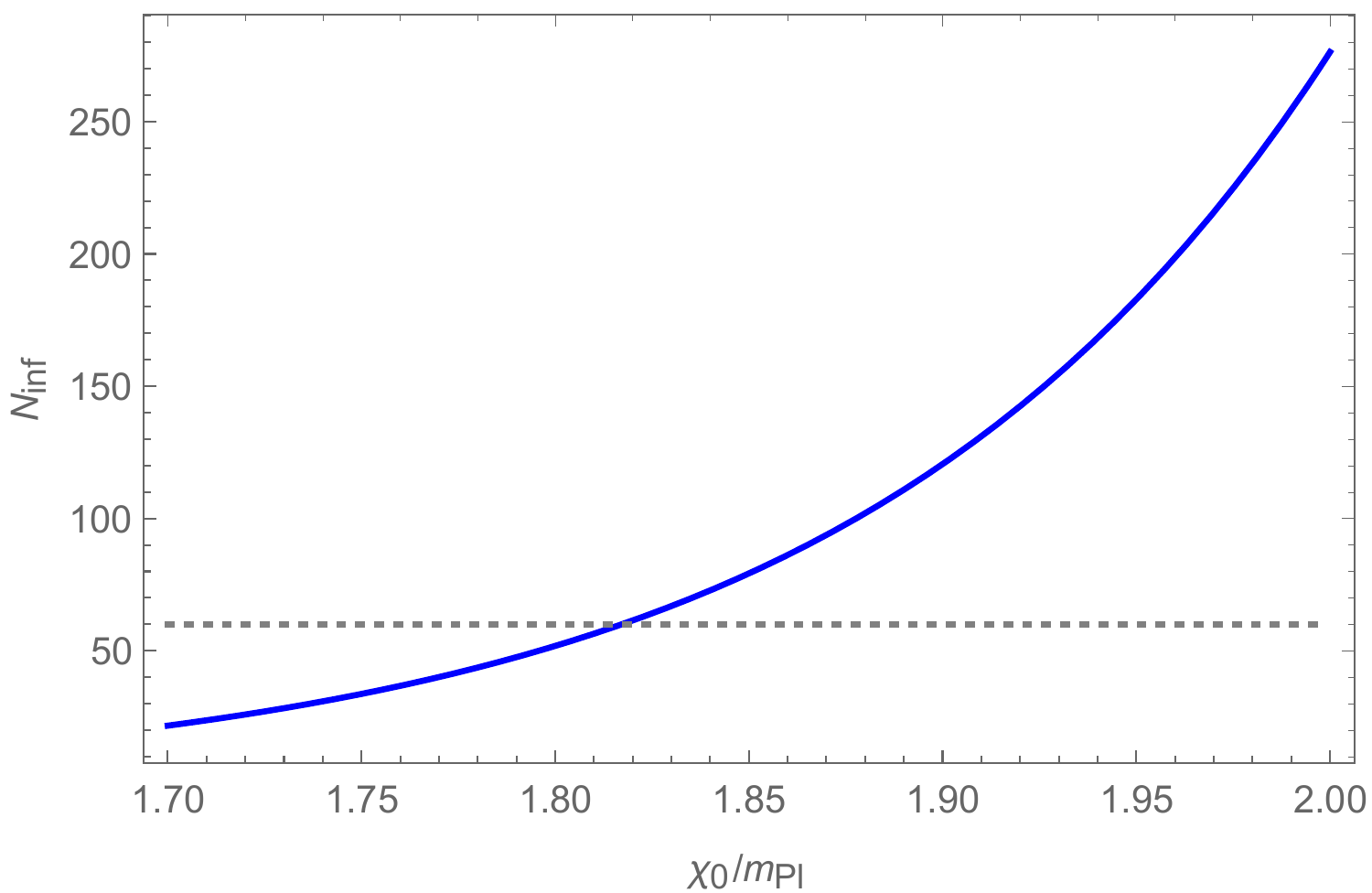}
  \caption{The $e$-folds $N_{\rm inf}$ during the slow-roll inflation for Starobinsky inflation ($\beta=3$) as a function of initial values of $\chi_0$ for the $\dot \chi_0 <0$. } 
  \label{efold} 
\end{figure}

Let us first consider the Starobinsky inflation in the framework of loop quantum BD cosmology, which corresponds to $\beta=3$. The Starobinsky inflation in LQC of GR has been discussed in \cite{zhu_preinflationary_2017a, bonga_phenomenological_2016}, in which both the dynamics of background and cosmological perturbations of Starobinsky inflation model have been extensivelly studied. One of main conclusions of these studies is that following the quantum bounce in LQC of GR, a desired slow-roll inflation phase is almost inevitable and the imprints of quantum bounce on primordial perturbation spectra and non-Gaussianities can be well within observational constraints \cite{zhu_preinflationary_2017a}. However, in loop quantum BD cosmology, as we mentioned, the Universe is not starting at the bounce. Thus it is interesting to explore the dynamics of the Starobinsky inflation in the framework of loop quantum BD cosmology and compare their differences with that in LQC of GR.

At the initial time, the Universe is expanding when the initial velocity is positive ($\dot \chi_0 >0$) and contracting if it is negative ($\dot \chi_0<0$). Thus our numerical analysis shall consider these two cases separately by paying particular attention to two important issues, namely how likely the occurrence of the slow-roll inflation is, and whether enough $e$-folds can be generated during the slow-roll inflation. 

For $\dot \chi_0 >0$, the results of the background evolution are illustrated in Fig.~\ref{dot_chi>0}. For the evolution of the scale factor (shown in the Top panel of Fig.~\ref{dot_chi>0}), it is shown clearly that the Universe is initial expanding from a finite Universe at $\hat t_0$.  During this expanding phase, the velocity $\dot \chi$ is decreasing so is the Hubble parameter $\hat H$. Therefore the expanding of Universe slows down until the Universe reaches its local maximum value and then collapses into a contracting phase. This picture is dramatically different from that in LQC of GR where the Universe is at the expanding phase right after the quantum bounce. After the contracting phase, the Hubble parameter again approaches zero and the bounce occurs, finally the Universe enters into the expanding phase (hereafter we use $\hat t_{\rm B}$ to denote the bounce point). During this pre-inflationary quantum phase, it is shown in the middle panel of Fig.~\ref{dot_chi>0} that the quantity $r_{\chi}$ is very close to unity. This means that the dynamics are dominated by the quantum geometry effects of LQBDC over this phase. The very different behaviors of the evolutions between the above phase and that in LQC of GR originate essentially from the loop quantization in the two different frames. 

Right after the bounce point $\hat t_{\rm B}$, the quantity $r_{\chi}$ quickly decreases to zero and therefore the Universe soon enters into the classical regime in which the quantum geometry effects are negligible and the evolution of the Universe follows equations in the classical BD cosmology. Now an essential question is whether the slow-roll inflation occurs after the above mentioned quantum processes. From the bottom panel of Fig.~\ref{dot_chi>0}, we see clearly that the slow-roll parameter $\hat \epsilon_{H}$ reduces quickly to a very small value ($\hat \epsilon_H \ll 1$) after the quantum gravity regime. This phase exactly represents the slow-roll inflation and the scale factor $\hat a(\hat t)$ is exponentially growing, as shown in the first panel of Fig.~\ref{dot_chi>0}. Further numerical analysis for more initial conditions show that the corresponding $e$-folds produced during the slow-roll inflation is sufficient to be larger than $60$ for any values of $\chi_{0}$ in the range of $(-\infty, \chi_{\rm max})$, as shown in Table.~\ref{smalltable1}. In Table.~\ref{smalltable1}, we also present the results from the numerical analysis for different values of $\chi_0$ for $\dot \chi_0>0$. 

For $\dot \chi_0 <0$, in contrast to the case of $\dot \chi_0 >0$, the Universe is initially contracting. The background evolutions for a set of initial conditions with $\dot \chi_0 <0$ is illustrated in Fig.~\ref{dot_chi<0} and Table.~\ref{smalltable1}, in which the scale factor $\hat a (\hat t)$, $r_{\chi}$, and the slow-roll parameter $\hat \epsilon_{H}$ are all obtained numerically. It is shown from the top panel of Fig.~\ref{dot_chi<0} that after the initial contracting phase dominated by the quantum geometry effects, the universe bounces to the expanding phase, during which the Universe eventually evolves into the slow-roll inflation. The corresponding $e$-folds $N_{\rm inf}$ during the slow-roll inflation as a function of $\chi_{0}$ is presented in Fig.~\ref{efold}. In order to produce sufficient $60$ $e$-folds during the slow-roll inflation, it is shown clearly that one has to require 
\bqn
\chi_0 \in (1.82 m_{\rm Pl},\; 2.32 m_{\rm Pl}).
\eqn
Another property of $N_{\rm inf}$ is that it increases as the value of $\chi_0$ increases until it reaches a maximum value when $\chi_0$ approaches its up bound. 

Here we would like to provide a brief summary about the background evolution of Starobinsky inflation. We find for both the initial positive and negative velocity, the evolution of the background can be in general divided into three different phases: {\em the pre-inflationary quantum phase, quantum-to-classical transition, and slow-roll inflationary phase}. During {\em pre-inflationary quantum phase}, the evolution of the background is dominated by the quantum effects of the loop quantum BD cosmology because $r_{\chi} \simeq 1$. It is shown that the quantum bounce occurs no matter the Universe is initially expanding (for $\dot \chi_0>0$) or contracting (for $\dot \chi_0 <0$). For initial expanding Universe ($\dot \chi_0>0$), the Universe shall first collapse to a contracting phase, and then bounce to a expanding phase, while for initial contracting Universe ($\dot \chi_0<0$), the Universe shall directly evolve to the expanding phase through the quantum bounce. For the {\em quantum-to-classical transition}, the quantity $r_{\chi}$ suddenly decreases from $r_{\chi} \simeq 1$ to $r_{\chi} \simeq 0$. Since $r_\chi$ denotes the ratio between the energy density of BD field and $\rho_{\rm c}$, this phase represents the intermediate region between the quantum and classical cosmology. Following this transition phase is the slow-roll inflationary phase and it is shown that for $\chi_0$ in restricted ranges the slow-roll inflation can lead to sufficient $60$ $e$-folds.

\begin{figure}
\centering
\includegraphics[width=8.5cm]{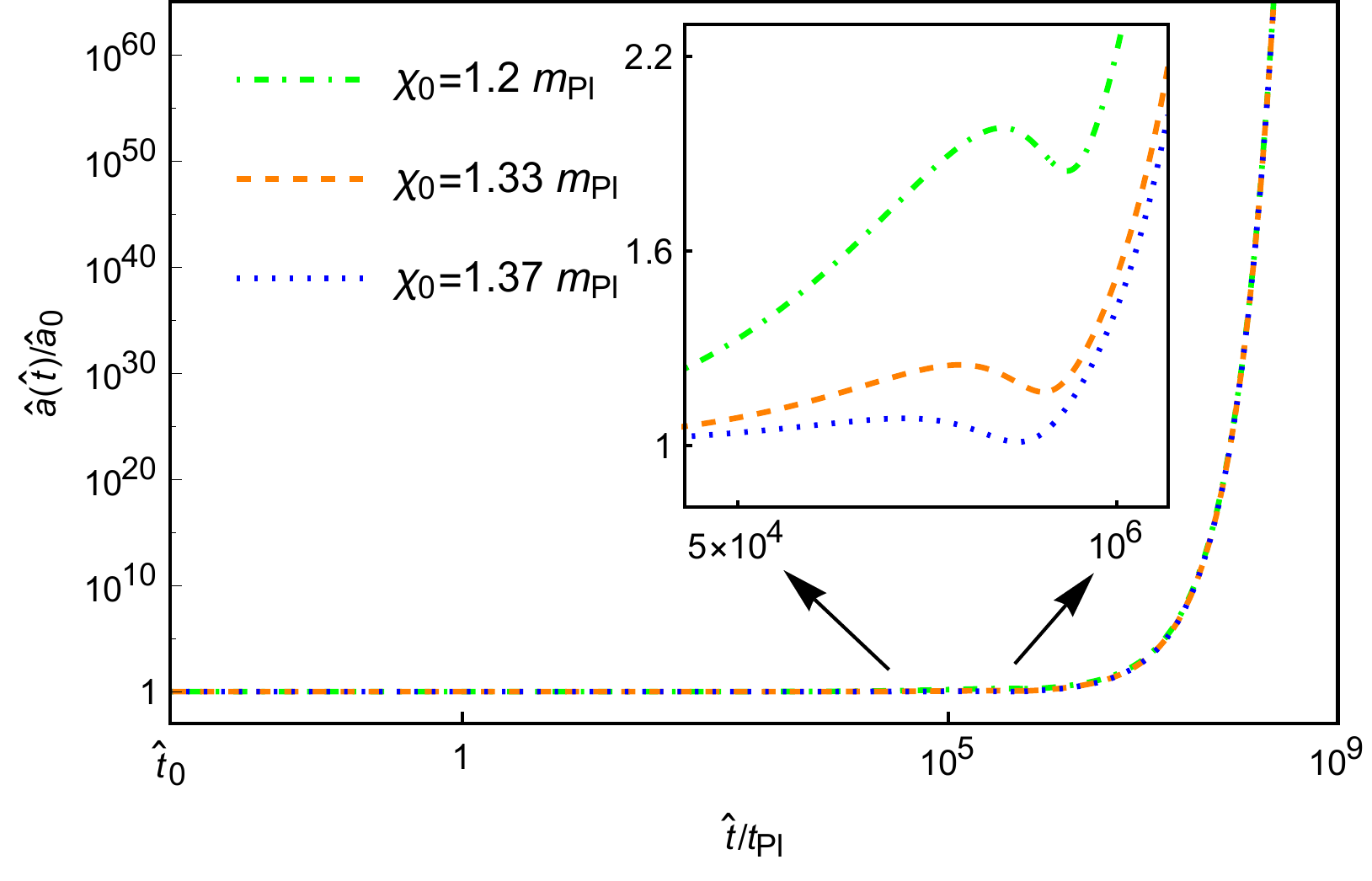}
\includegraphics[width=8.5cm]{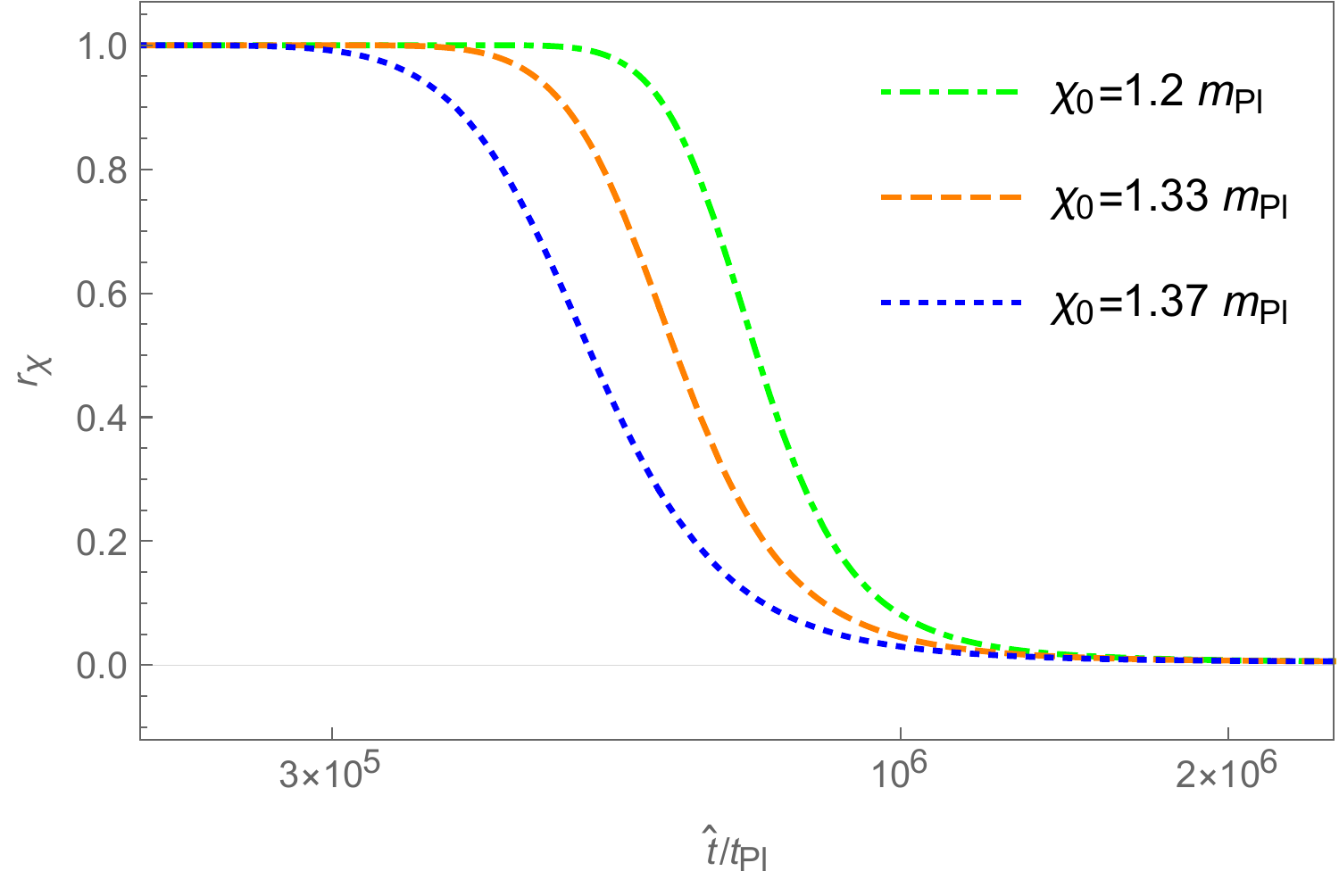}
\includegraphics[width=8.5cm]{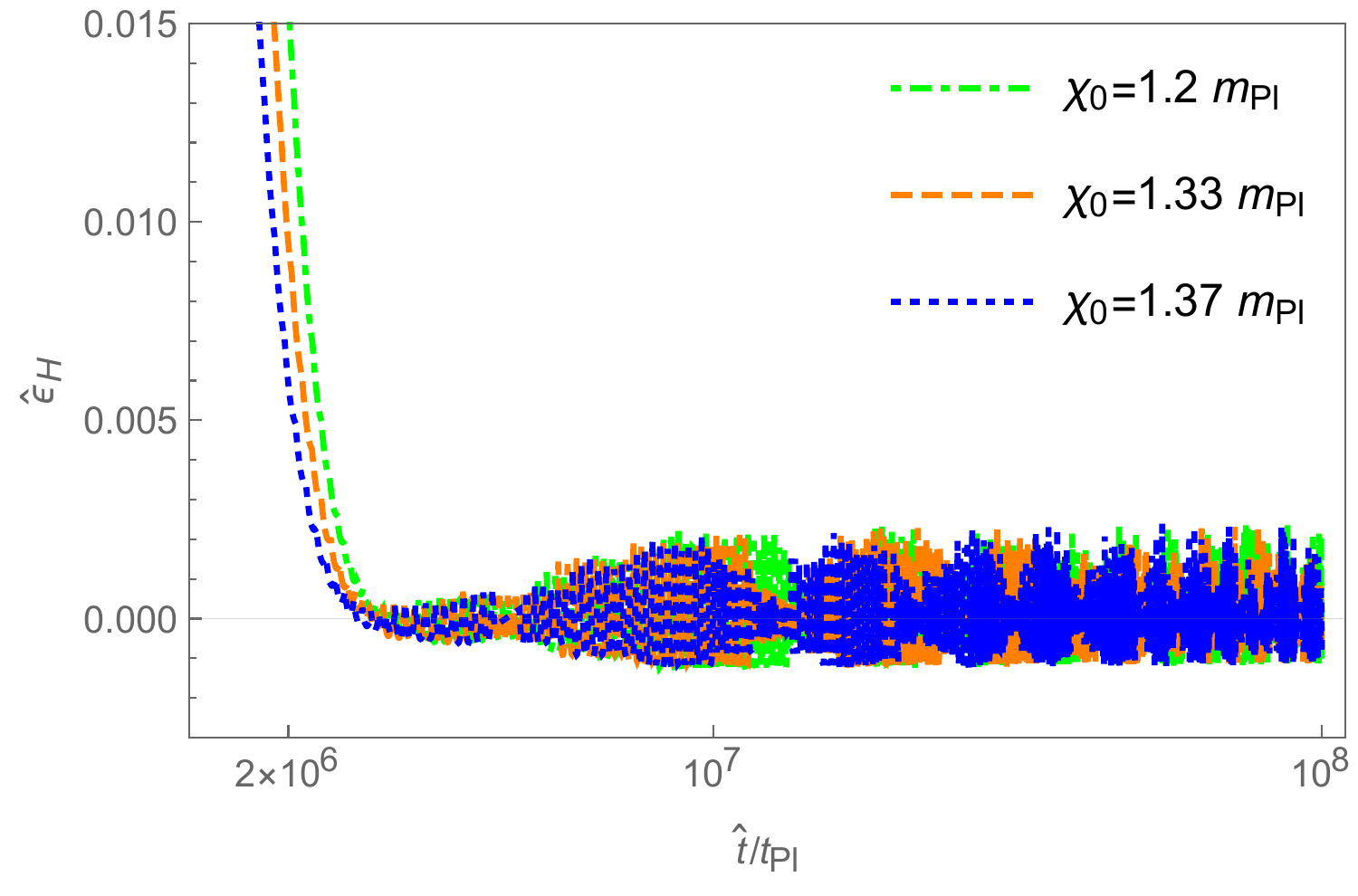}
\caption{The figure shows the numerical evolution of the scale factor $\hat a(\hat t)$, $r_{\chi}$ and the slow-roll parameter $\hat \epsilon_{H} $ for $\alpha$-attractor inflation with $\beta=1$ for $\dot \chi_0 > 0 $. }
\label{dot_chi1>0}
\end{figure}

\begin{figure}
\centering
\includegraphics[width=8.5cm]{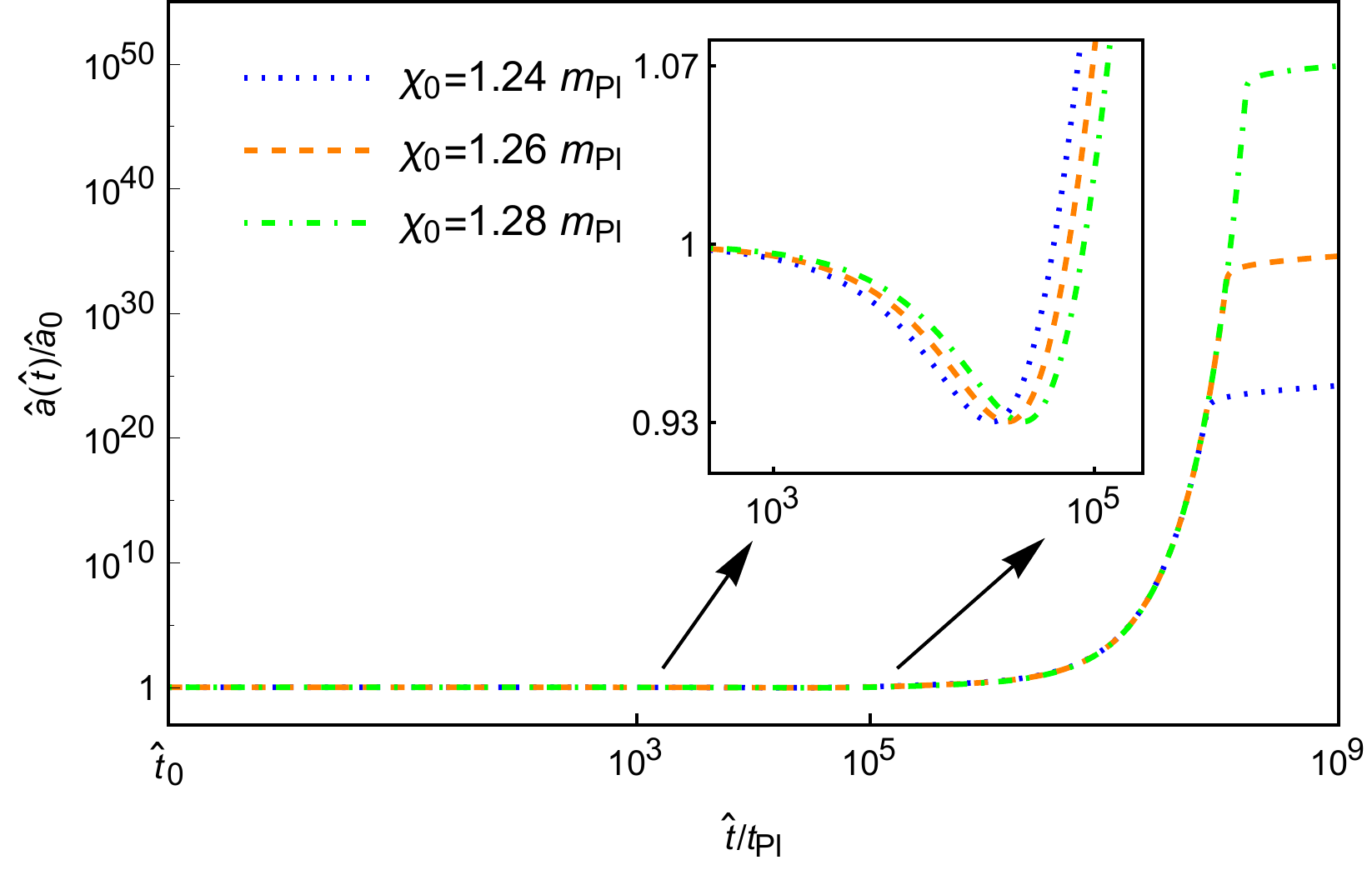}
\includegraphics[width=8.5cm]{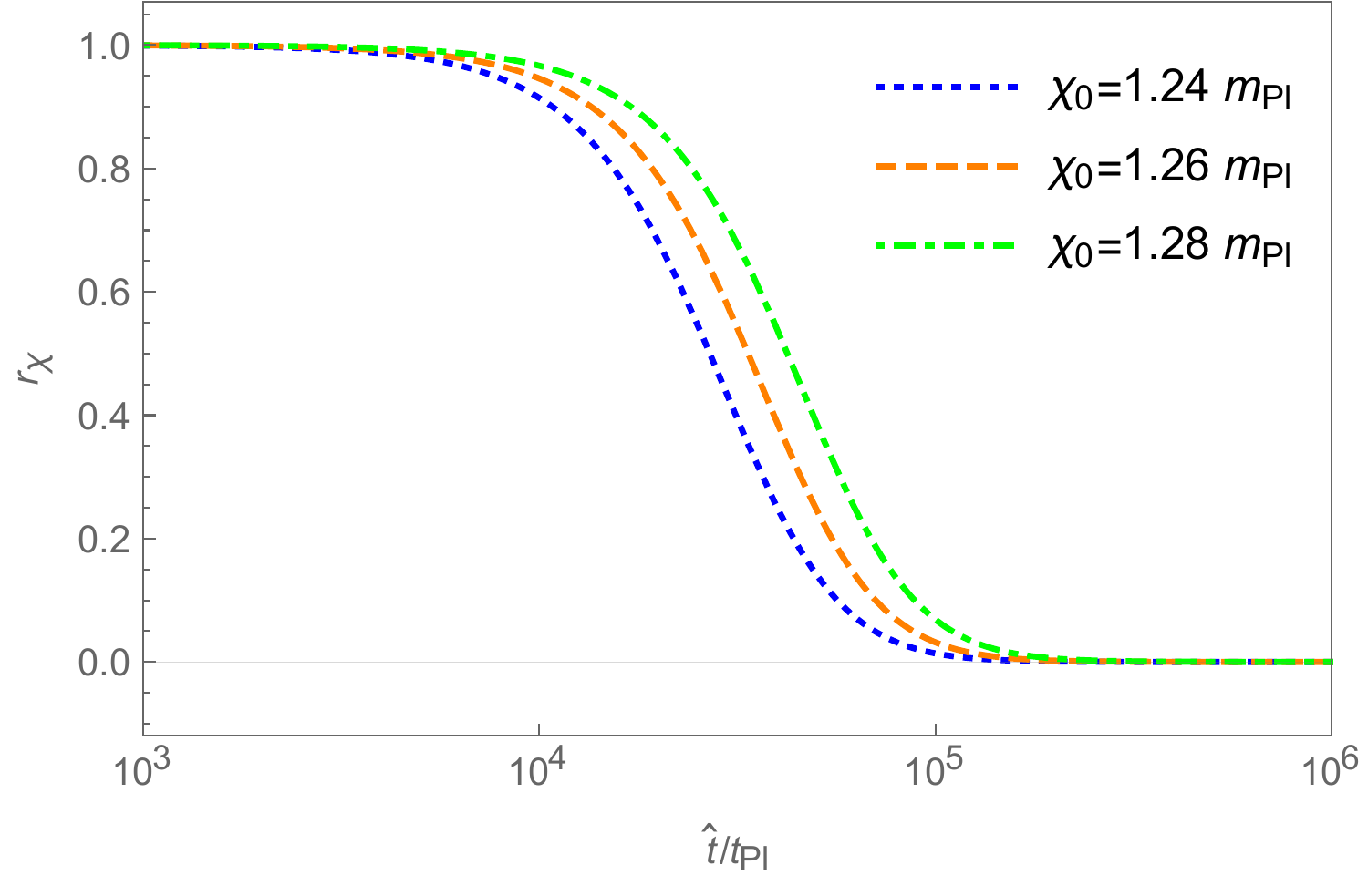}
\includegraphics[width=8.5cm]{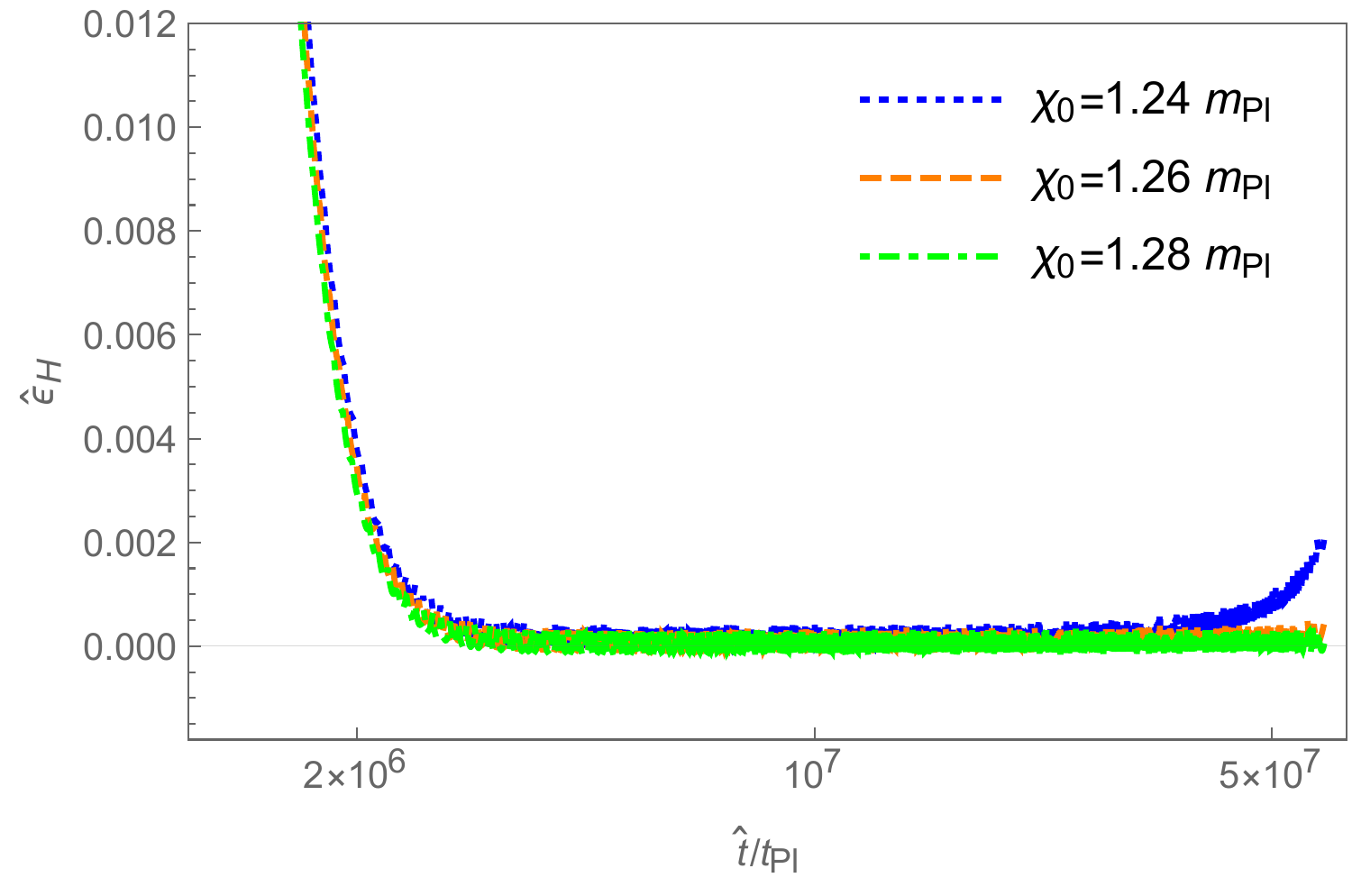}
\caption{The figure shows the numerical evolution of the scale factor $\hat a(\hat t)$, $r_{\chi}$ and the slow-roll parameter $\hat \epsilon_{H} $ for $\alpha$-attractor inflation with $\beta=1$ for $\dot \chi_0 < 0 $. }
\label{dot_chi1<0}
\end{figure}

 \begin{figure}
  \centering
  \includegraphics[width=8.5cm]{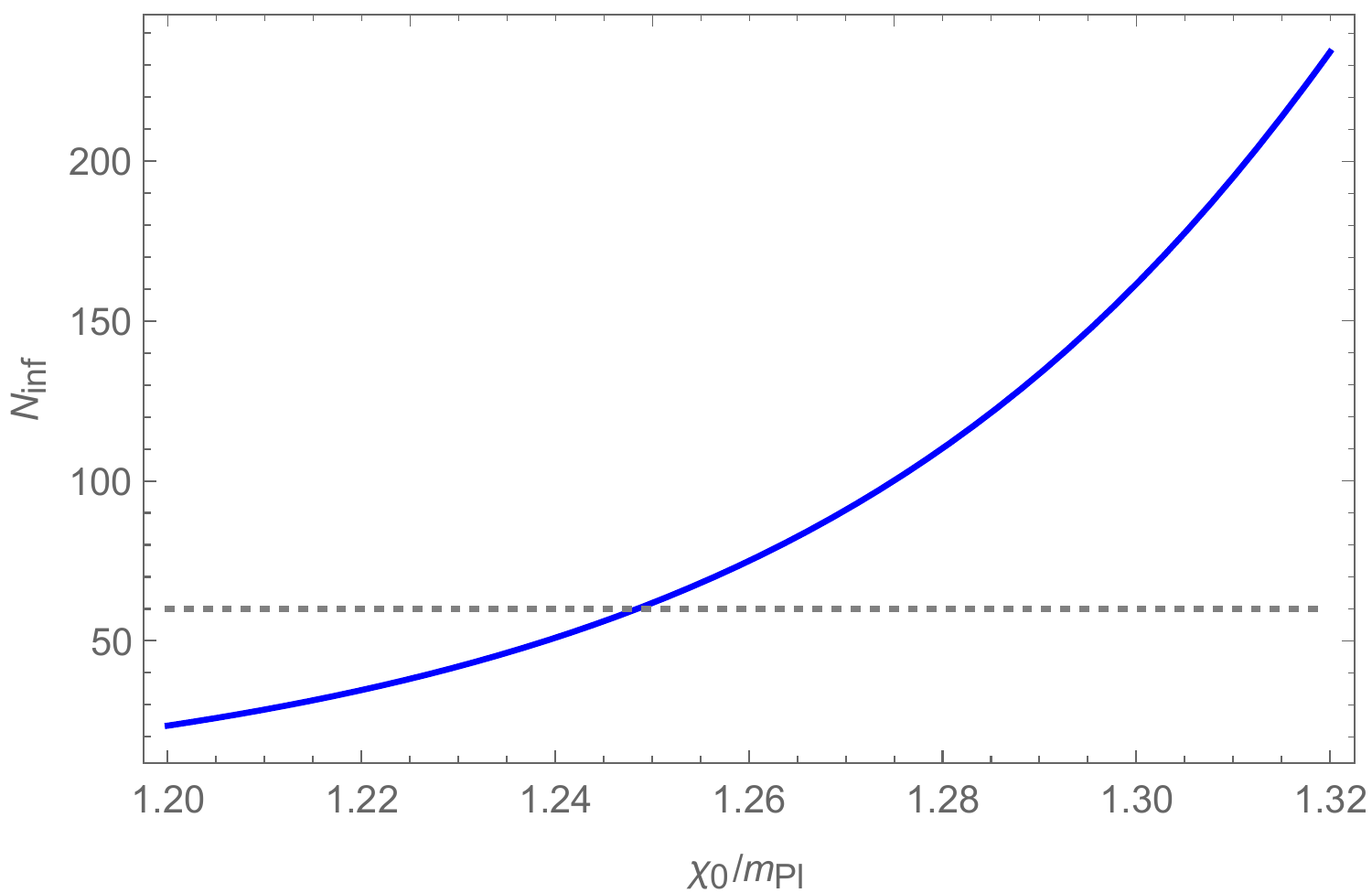}
  \includegraphics[width=8.5cm]{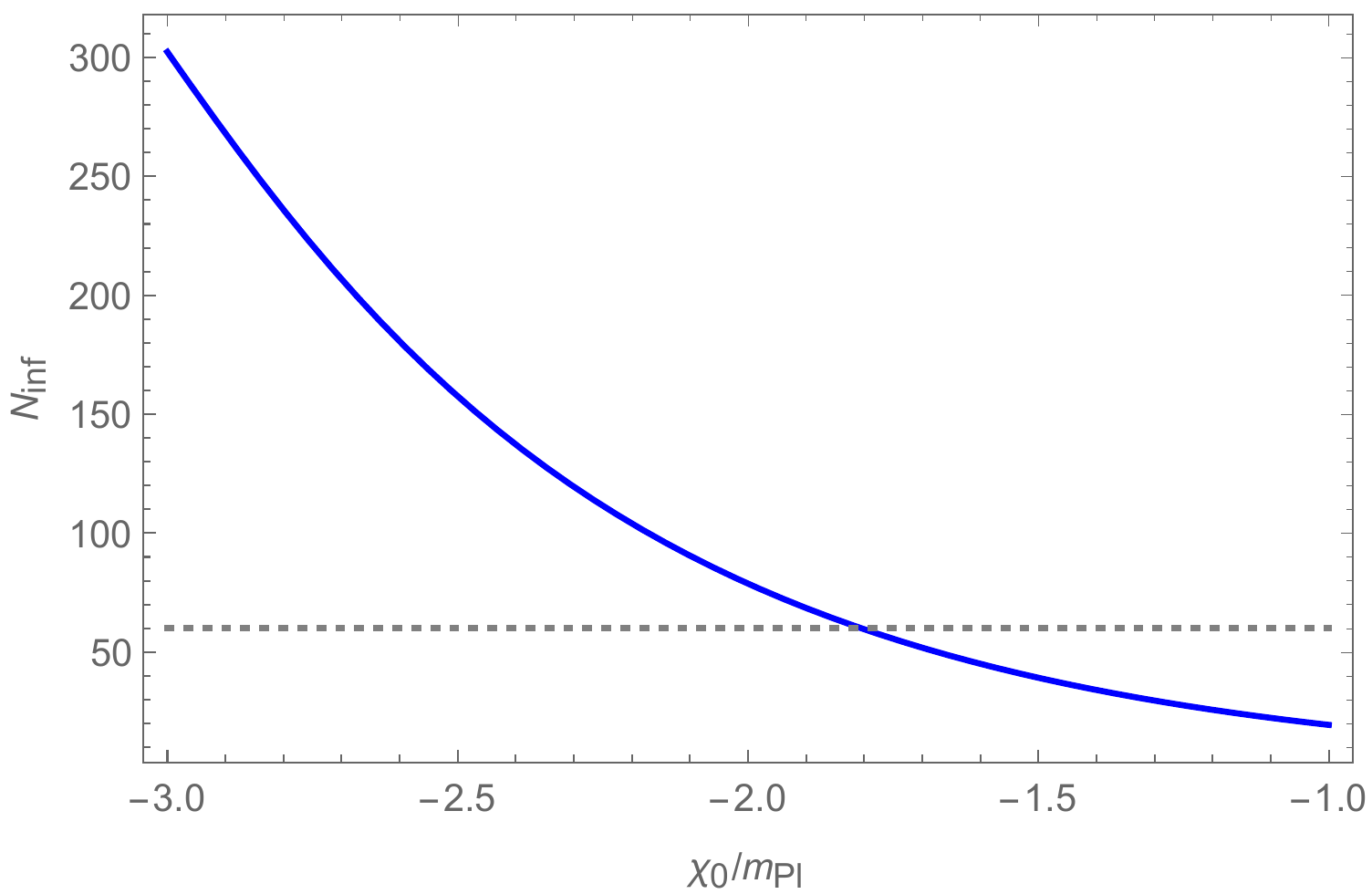}
  \caption{For $\beta=1$, the e-folds $N_{\rm inf}$ during the slow-roll inflation as a function of $\chi_0$ with $\dot \chi_0<0$. Top panel: for positive values of $\chi_0$. Bottom panel: for negative values of $\chi_0$.} 
  \label{beta1efold} 
\end{figure}

\begin{table*}[!h]
%	\centering
	\caption{Table for the cases $\beta=$3. The value of $\hat t_{B}$ denotes the time when a quantum bounce occurs. We also define $\hat t_{C}$ as the time when $r_{\chi}$ decreases below $10^{-3}$ and consider the evolution equations classically after $\hat t=\hat t_{C}$ .} \lb{smalltable1}
\begin{tabular}{p{0.09\textwidth}<{\centering}p{0.11\textwidth}<{\centering}p{0.12\textwidth}<{\centering}p{0.13\textwidth}<{\centering}p{0.12\textwidth}<{\centering}p{0.13\textwidth}<{\centering}p{0.12\textwidth}<{\centering}p{0.12\textwidth}<{\centering}}
		\hline
\specialrule{0.04em}{2.2pt}{3pt}
$\dot \chi_0 $   &  $\chi_0$   & $\hat t_B/t_{\rm Pl}$   &   $\hat t_C/t_{\rm Pl}$   & Inflation  & $\hat t/t_{\rm Pl}$        & $\chi_*$         & $N_{\rm inf}$ \\ \specialrule{0.04em}{3pt}{3pt}
\multirow{2}{*}{positive}& \multirow{2}{*}{2.3}  & \multirow{2}{*}{$ 3.7\times 10^5$}& \multirow{2}{*}{$ 2.1\times 10^8$}  &   starts    &     $ 2.8\times 10^5$     & 2.31              & \multirow{2}{*}{549.67}  \\
          &                         &                                       &                                     &  ends      &     $ 9.8\times 10^8$     &0.12                &        \\ \specialrule{0em}{3pt}{3pt}
	     & \multirow{2}{*}{2}   &  \multirow{2}{*}{$ 6.3\times 10^5$}     & \multirow{2}{*}{$ 2.1\times 10^8$}   &   starts    &     $ 5.3\times 10^5$     & 2.31               & \multirow{2}{*}{544.97}  \\
          &                         &                                      &                                      &  ends      &     $ 9.8\times 10^8$     &0.12                &       \\  \specialrule{0em}{3pt}{3pt}
	     & \multirow{2}{*}{0}   &  \multirow{2}{*}{$ 6.7\times 10^5$}     & \multirow{2}{*}{$ 2.1\times 10^8$}   &   starts    &     $ 5.7\times 10^5$     & 2.31               & \multirow{2}{*}{544.97}  \\
          &                         &                                      &                                      &  ends      &     $ 9.8\times 10^8$     &0.12                &       \\  \specialrule{0em}{3pt}{3pt}
	     & \multirow{2}{*}{-2}   &  \multirow{2}{*}{$ 6.7\times 10^5$}     & \multirow{2}{*}{$ 2.1\times 10^8$}   &   starts    &     $ 5.7\times 10^5$     & 2.31               & \multirow{2}{*}{544.97}  \\
          &                         &                                      &                                      &  ends      &     $ 9.8\times 10^8$     &0.12                &       \\  \specialrule{0em}{3pt}{3pt}
	     & \multirow{2}{*}{-5}   &  \multirow{2}{*}{$ 6.7\times 10^5$}     & \multirow{2}{*}{$ 2.1\times 10^8$}   &   starts    &     $ 5.7\times 10^5$     & 2.31               & \multirow{2}{*}{544.97}  \\
          &                         &                                      &                                      &  ends      &     $ 9.8\times 10^8$     &0.12                &       \\  \specialrule{0em}{3pt}{3pt}
 \specialrule{0.04em}{4pt}{4pt}
\multirow{2}{*}{negative}& \multirow{2}{*}{2}   & \multirow{2}{*}{9810.43}      &  \multirow{2}{*}{$ 1.5\times 10^5$}  &   starts    &     $ 3.2\times 10^5$     & 1.55               & \multirow{2}{*}{276.3}  \\
          &                         &                                       &                                     &  ends      &     $ 2.4\times 10^8$     &0.12                &        \\ \specialrule{0em}{3pt}{3pt}
	     & \multirow{2}{*}{1.82}   &  \multirow{2}{*}{3228.85}            & \multirow{2}{*}{$ 4.7\times 10^4$}   &   starts    &     $ 3.3\times 10^5$     & 1.19               & \multirow{2}{*}{61.45}  \\
          &                         &                                      &                                      &  ends      &     $ 5.6\times 10^7$     &0.12                &       \\  \specialrule{0em}{3pt}{3pt}
	    & \multirow{2}{*}{1.6}     &  \multirow{2}{*}{836.80}             &  \multirow{2}{*}{$ 1.2\times 10^4$ } &   starts    &     $ 3.5\times 10^5$     & 0.75               & \multirow{2}{*}{8.58}  \\
          &                         &                                      &                                       &  ends      &     $ 9.2\times 10^6$     &0.12                &       \\  \specialrule{0em}{3pt}{3pt}
	     & \multirow{2}{*}{0}      &  \multirow{2}{*}{0.05}               &  \multirow{2}{*}{0.66}              &   starts    &                          &                    &       \\
          &                         &                                       &                                     &  ends      &                            &                    &       \\  \specialrule{0em}{3pt}{3pt}
	    & \multirow{2}{*}{-2}     & \multirow{2}{*}{$2.1\times 10^{-7}$} & \multirow{2}{*}{$ 3.0\times 10^{-6}$} &   starts    &                           &                   &      \\
          &                         &                                       &                                       &  ends      &                            &                  &       \\ \specialrule{0em}{3pt}{3pt}
\hline
\hline
\end{tabular}
\end{table*}

\subsection{$\alpha$-attractor inflation}

In this subsection, we begin to consider the $\alpha$-attractor inflation ($\beta \neq 3 $) and focus on several cases of $\beta=1,\; 5, \; 10, \; 20$. 

\subsubsection{$\beta=1$}

For $\beta=1$, the results of the background evolutions are illustrated with $\dot \chi_0 > 0 $ and $\dot \chi_0 < 0 $, respectively, in Figs.~\ref{dot_chi1>0} and \ref{dot_chi1<0}, in which the scale factor $\hat a(\hat t)$ , $r_{\chi} $ and the slow-roll parameter $\hat \epsilon_{H}$ are all obtained numerically. The behaviors of these solutions are very similar to the Starobinsky inflation. For those initial conditions that are able to realize the slow-roll inflation, the evolution of the universe can be divided into three phases: the pre-inflationary quantum phase, quantum-to-classcial transition, and the slow-roll inflationary phase. From Figs.~\ref{dot_chi1>0} and \ref{dot_chi1<0}, one can see that the behaviors of pre-inflationary quantum phase are almost the same as those of the Starobinsky inflation. 

For the slow-roll inflationary phase, let us first discuss the case of $\dot \chi_0 >0$. The numerical results for several initial conditions with $\dot \chi_0>0$ are presented in Table.~\ref{bigtable2}. We find that the desired slow-roll inflation can be produced for any values of $\chi_0$ in the range 
\bqn
\chi_0 \in (- \infty, \; 1.39 m_{\rm Pl}).
\eqn
For $\dot \chi_0 < 0$, the $e$-folds $N_{\rm inf}$ during the slow-roll inflation as a function of $\chi_0$ is illustrated in Fig.~\ref{beta1efold}, from which we can see that, in order to produce at least 60 $e$-folds during slow-roll inflation, the initial conditions have to be restricted to
\bqn
\chi_0 \in (-\infty ,-1.80m_{\rm Pl}) \cup (1.25 m_{\rm Pl}, 1.39 m_{\rm Pl}).
\eqn

\subsubsection{$\beta=5,10$ and $20 $}

\begin{figure*}
\centering
\includegraphics[width=5.8cm]{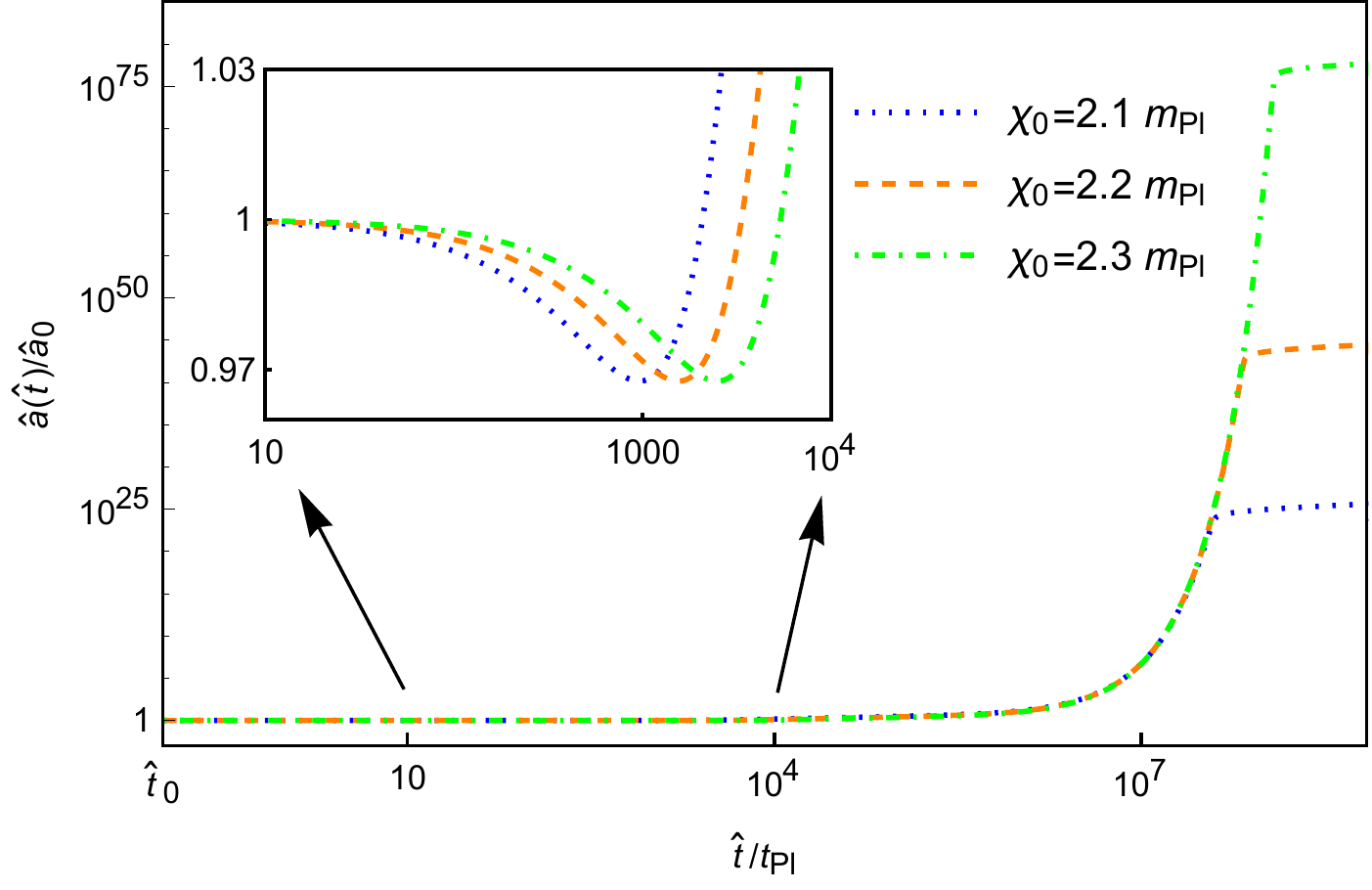}
\includegraphics[width=5.8cm]{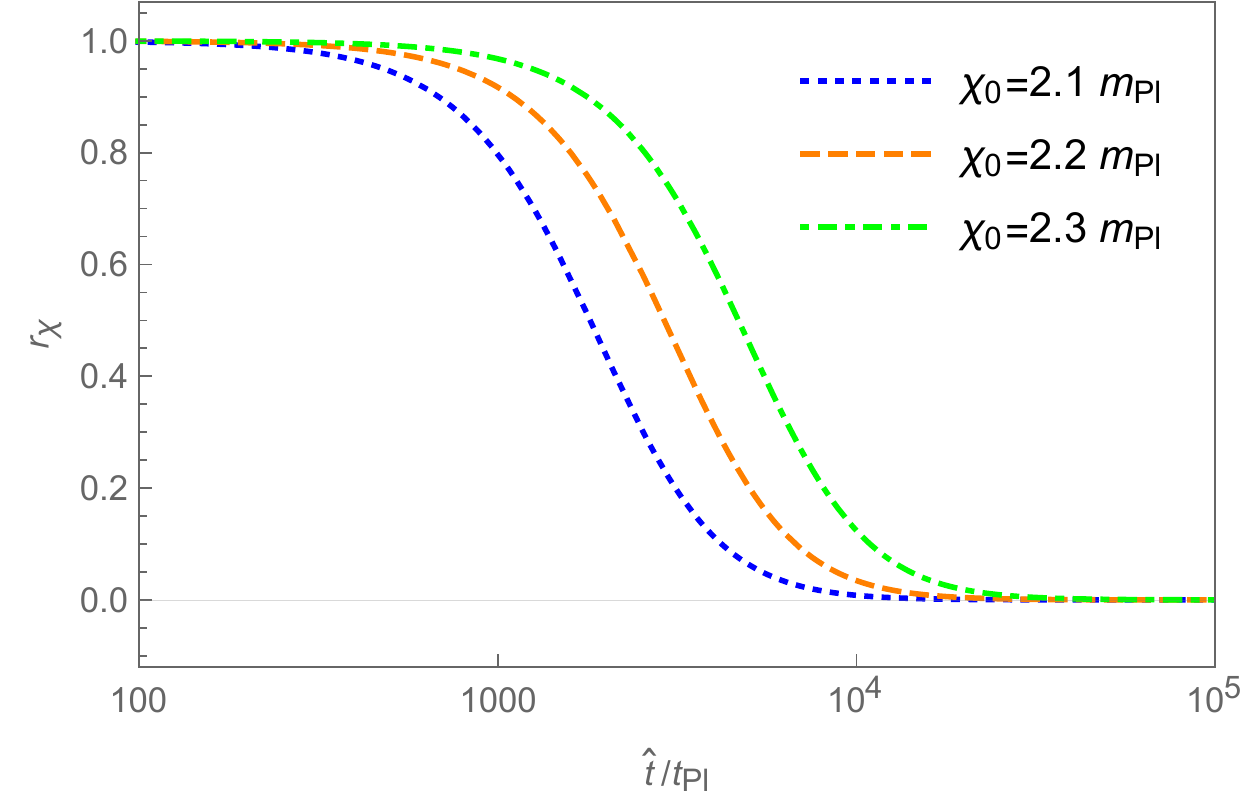}
\includegraphics[width=5.8cm]{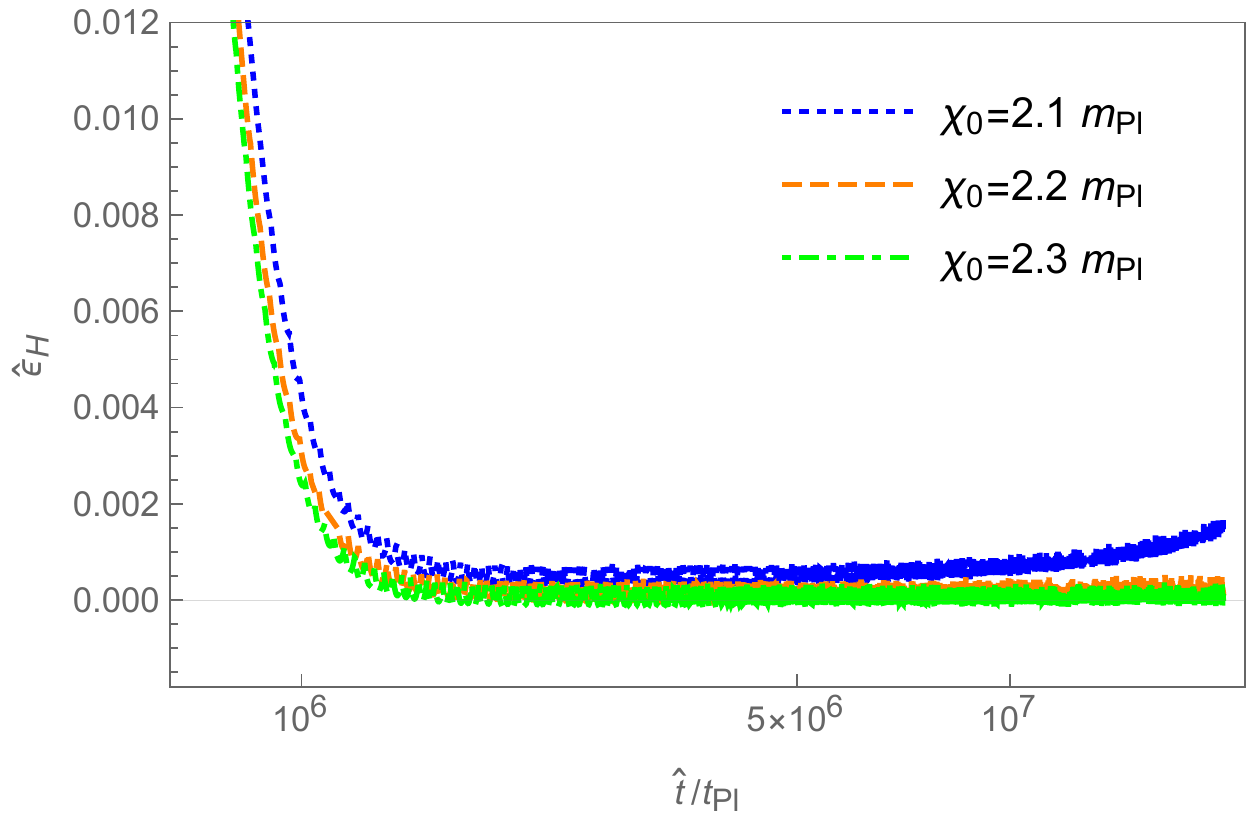}
\includegraphics[width=5.8cm]{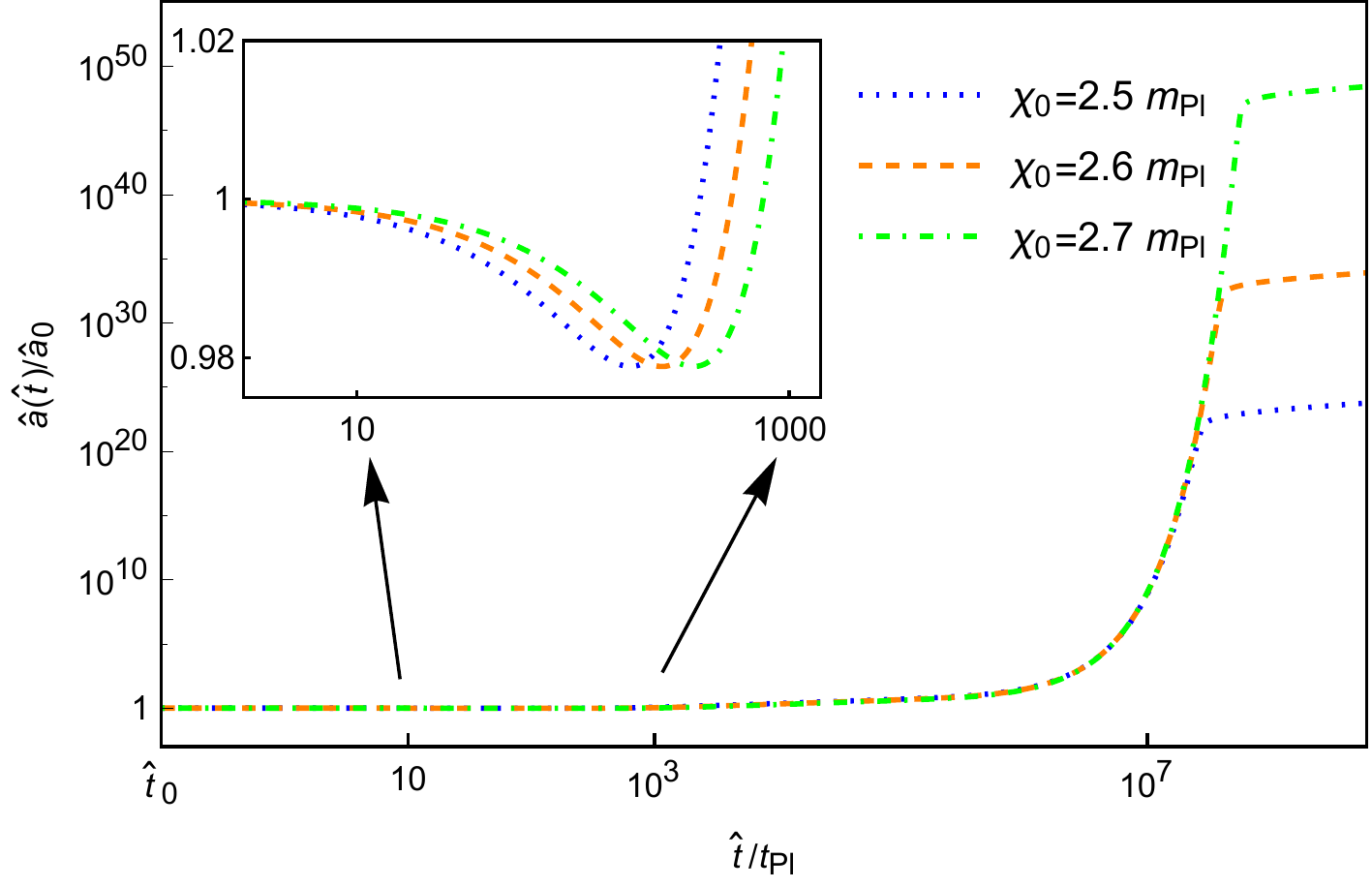}
\includegraphics[width=5.8cm]{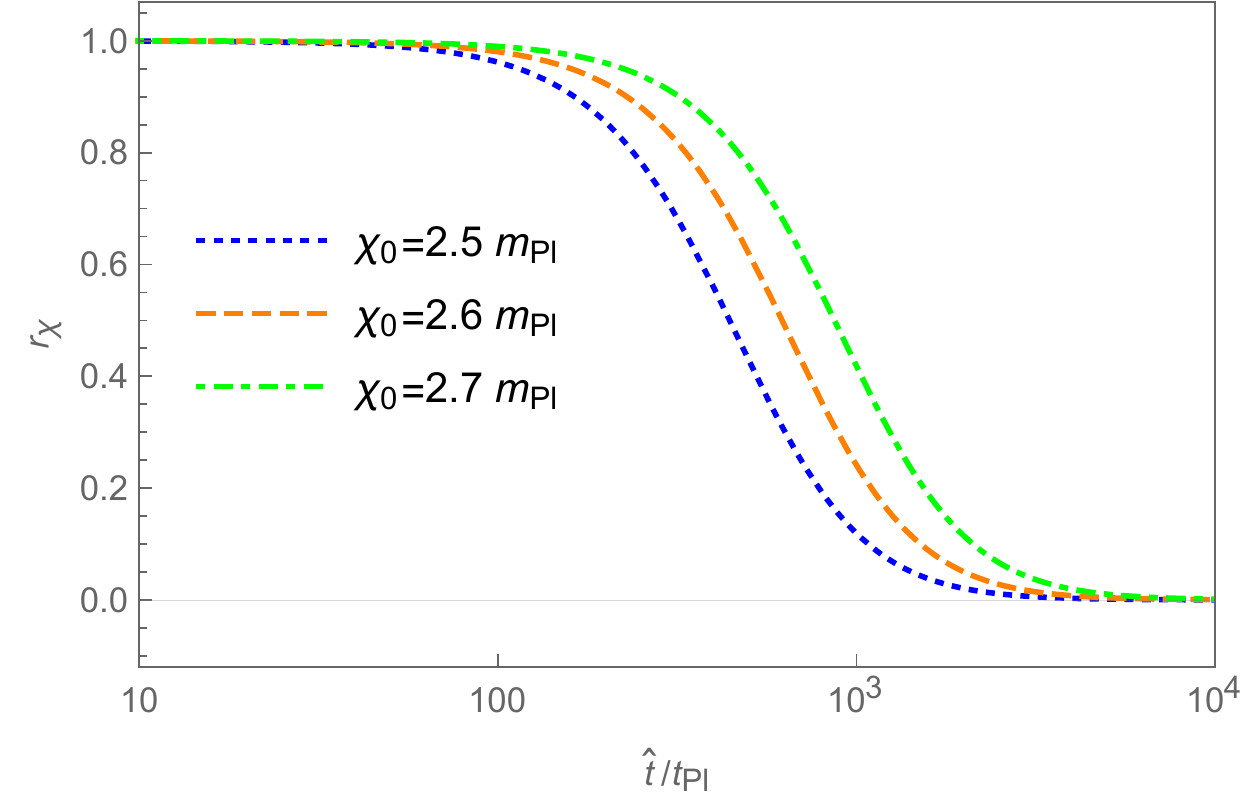}
\includegraphics[width=5.8cm]{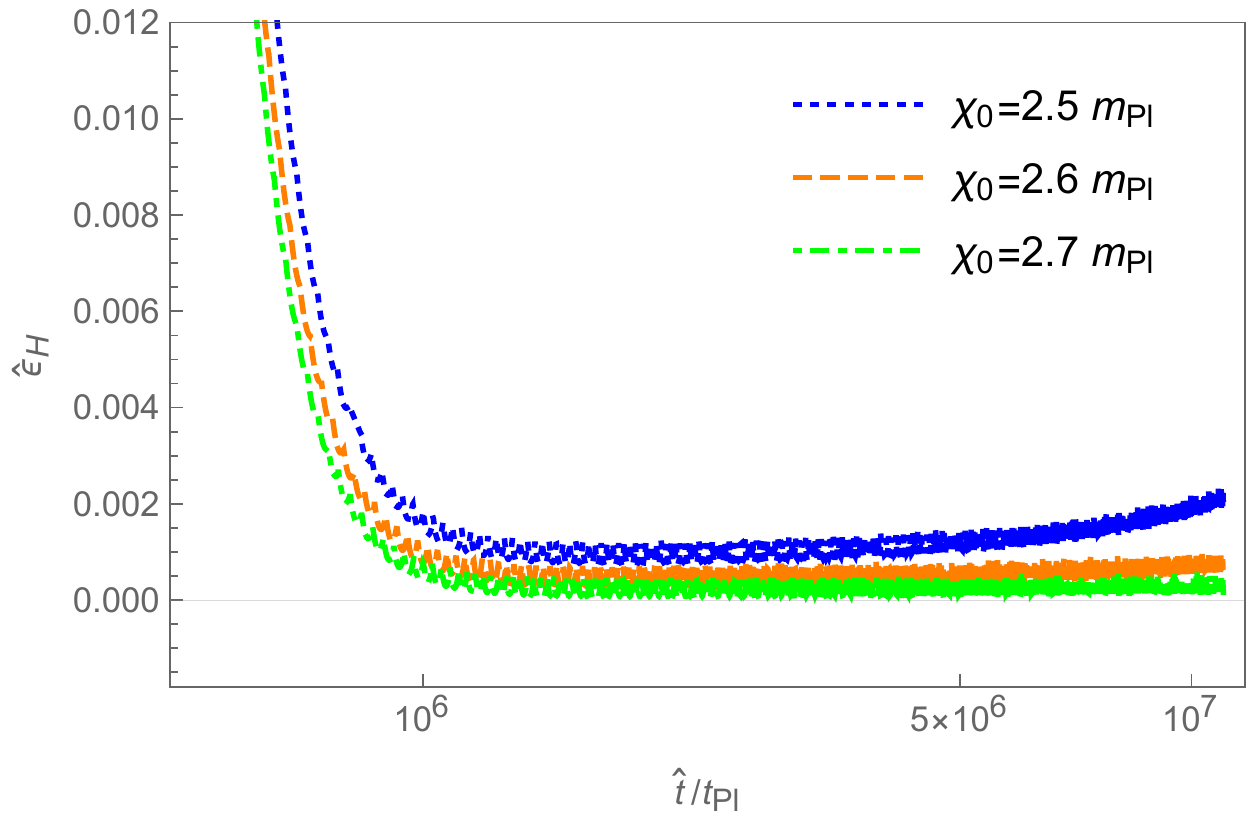}
\includegraphics[width=5.8cm]{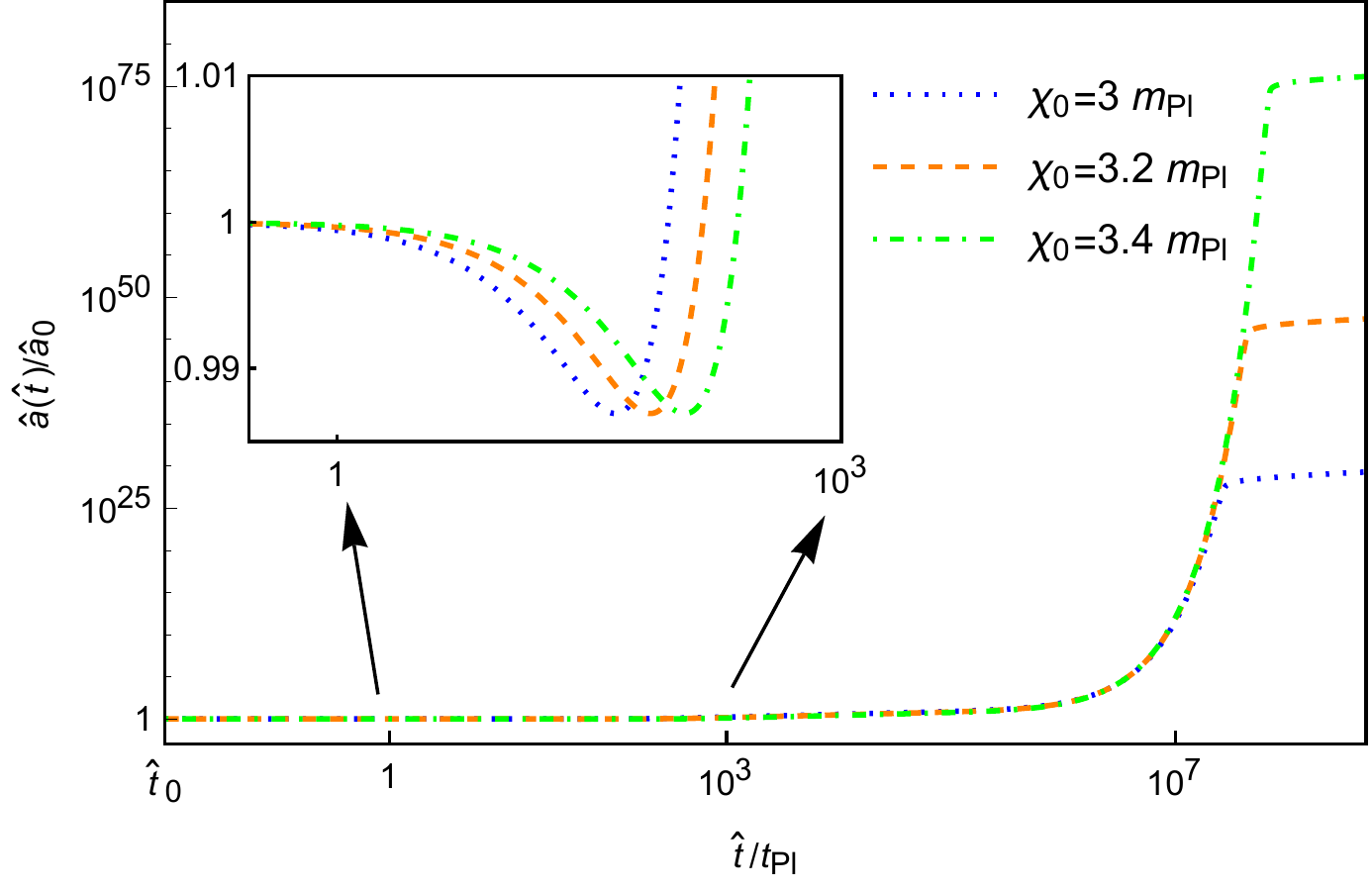}
\includegraphics[width=5.8cm]{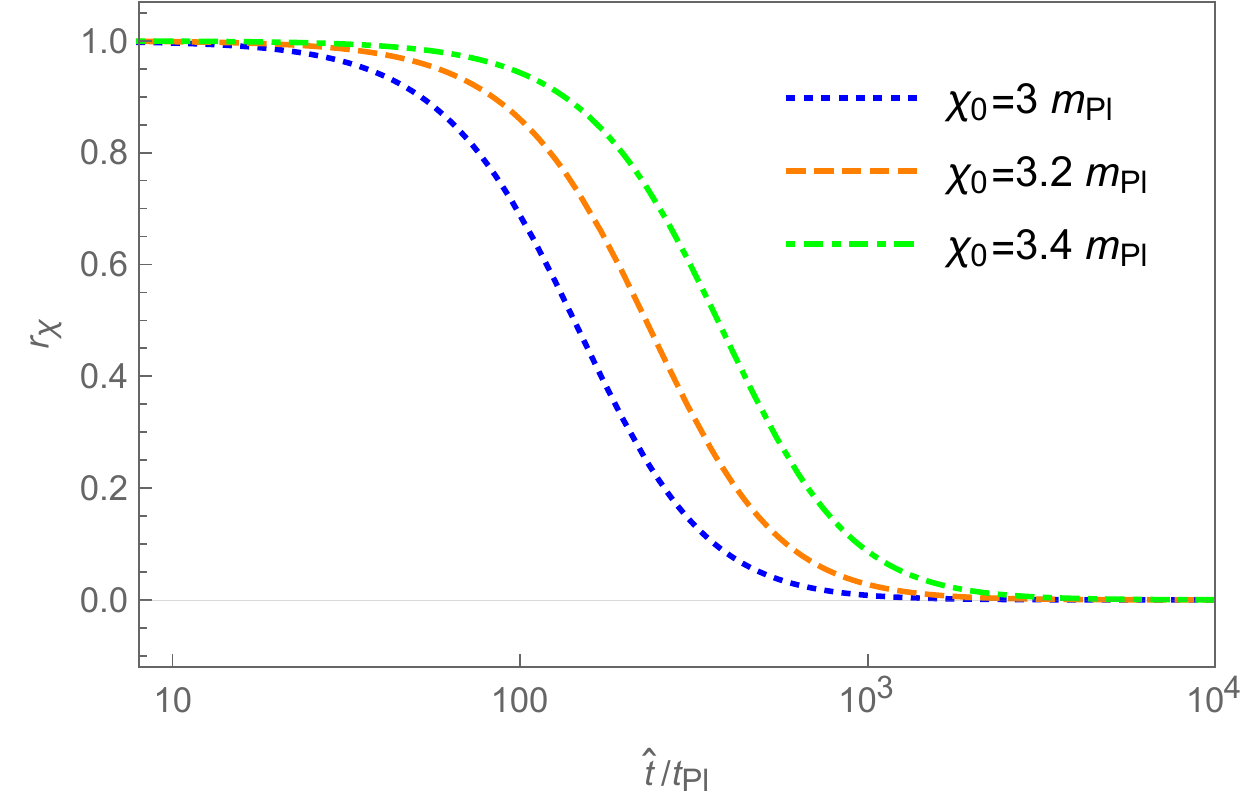}
\includegraphics[width=5.8cm]{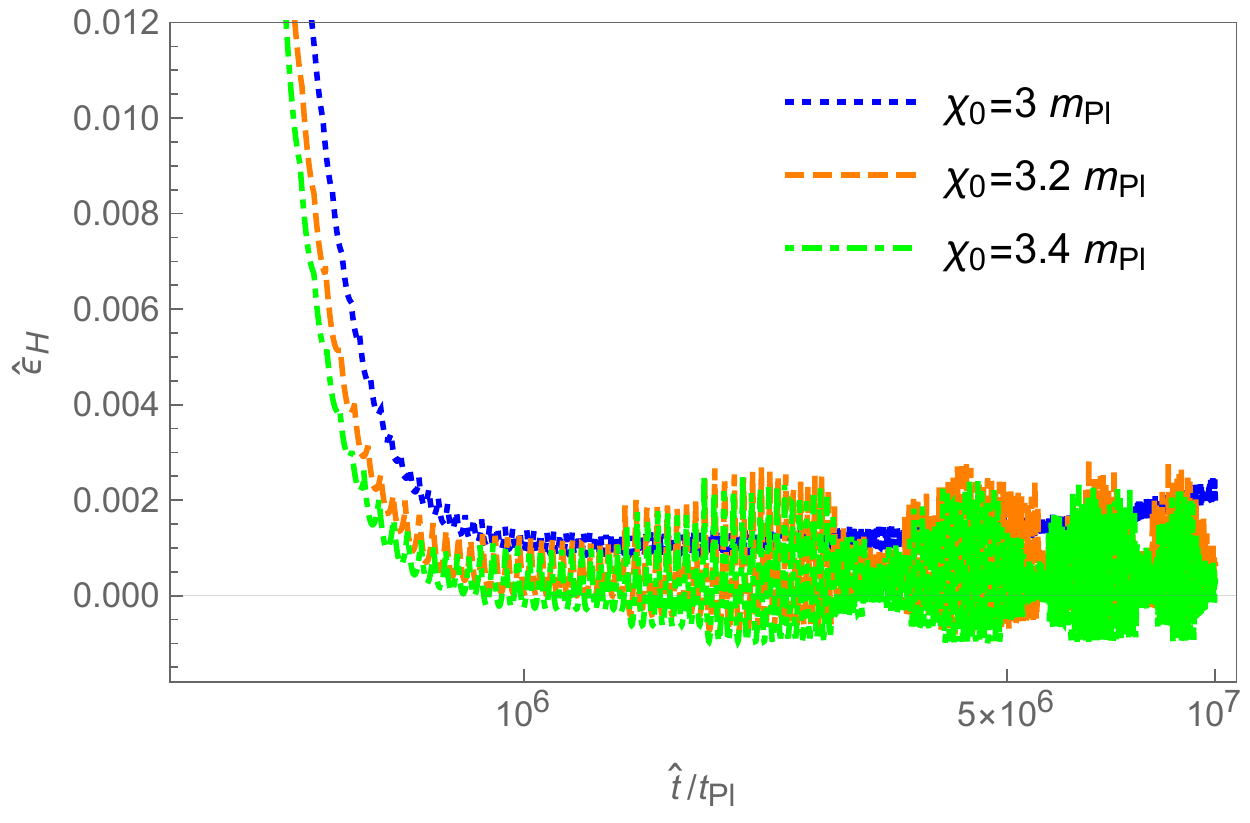}
\caption{The figure shows the numerical evolution of $\hat a(\hat t)$ ,$r_{\chi}$ and $\hat \epsilon_{H} $ with $\dot \chi_0 < 0 $. Top panel: $\beta=5$. Middle panel: $\beta=10$. Bottom panel: $\beta=20$. The insets show that a quantum bounce occurs.}
\label{bigfig1}
\end{figure*}

\begin{figure*}
\centering
\includegraphics[width=5.8cm]{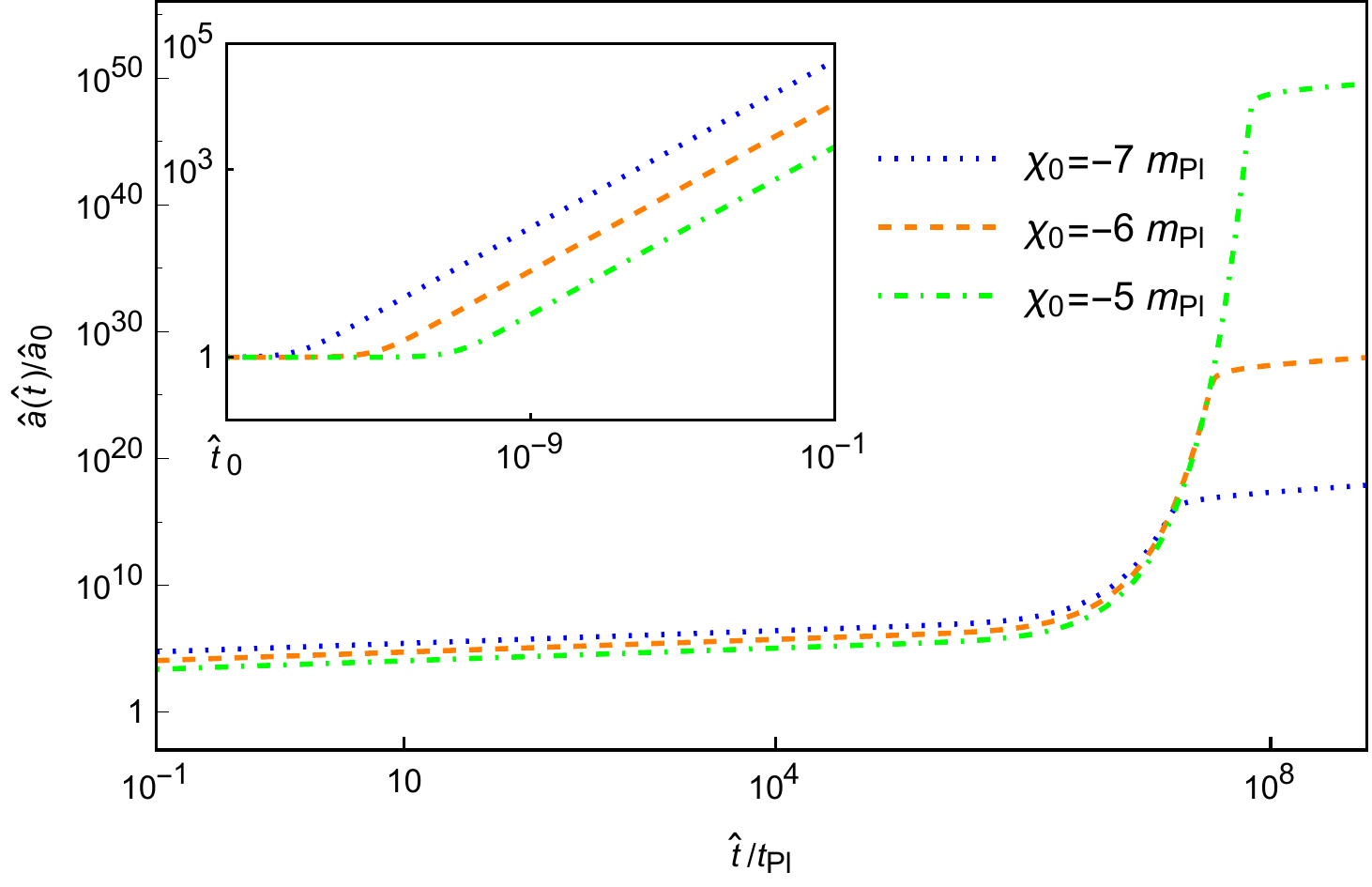}
\includegraphics[width=5.8cm]{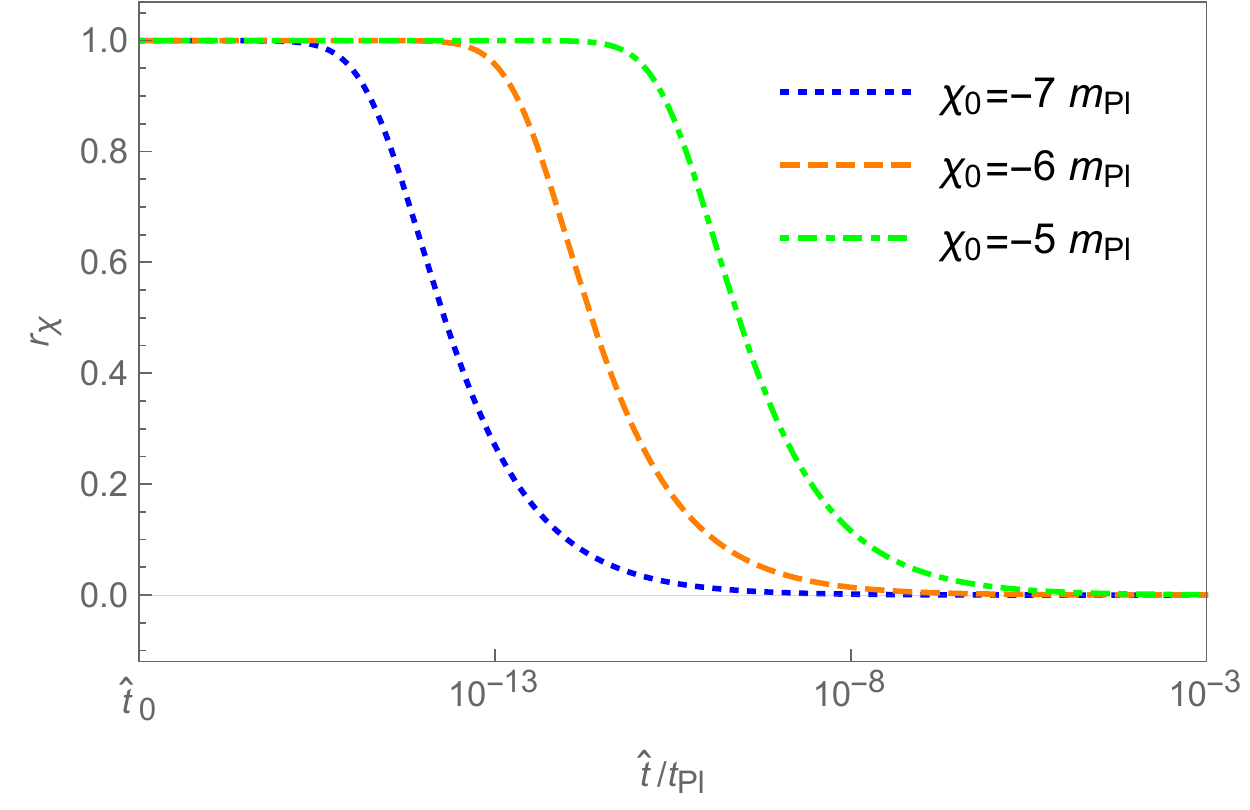}
\includegraphics[width=5.8cm]{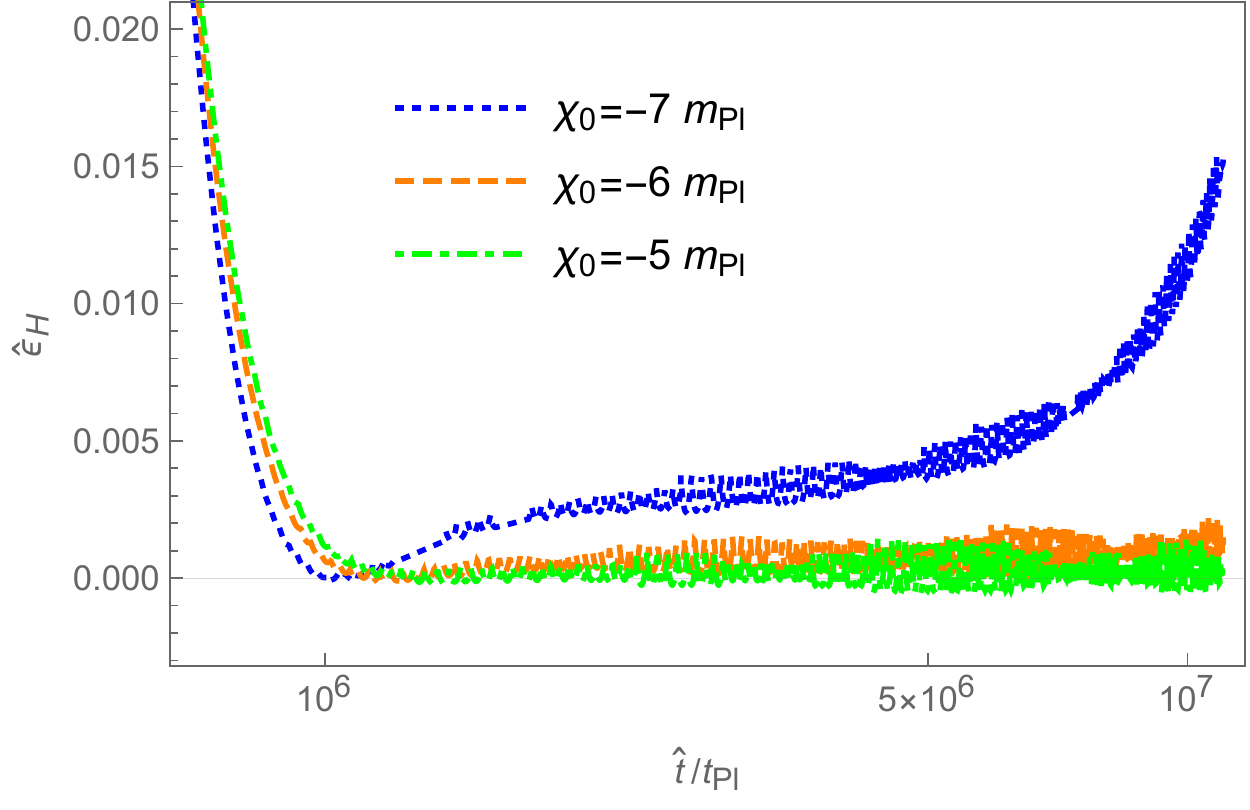}
\includegraphics[width=5.8cm]{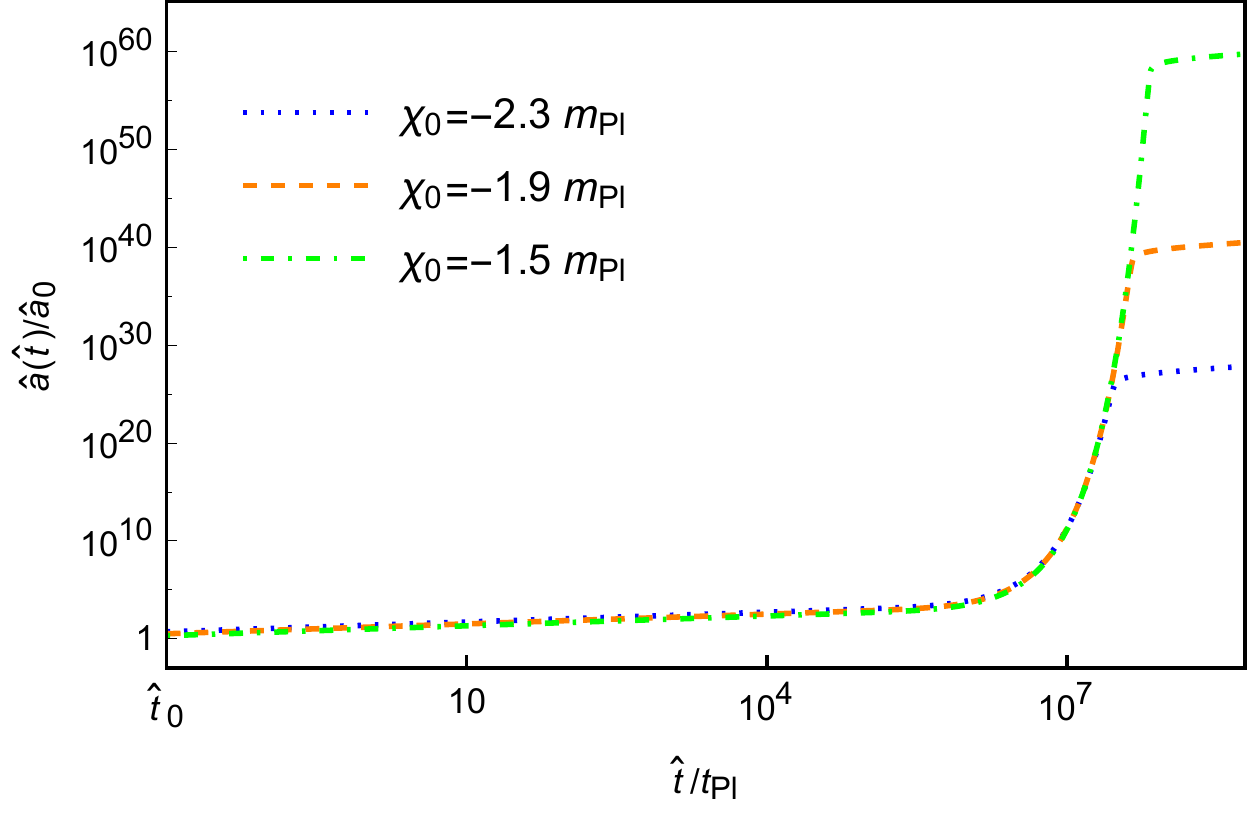}
\includegraphics[width=5.8cm]{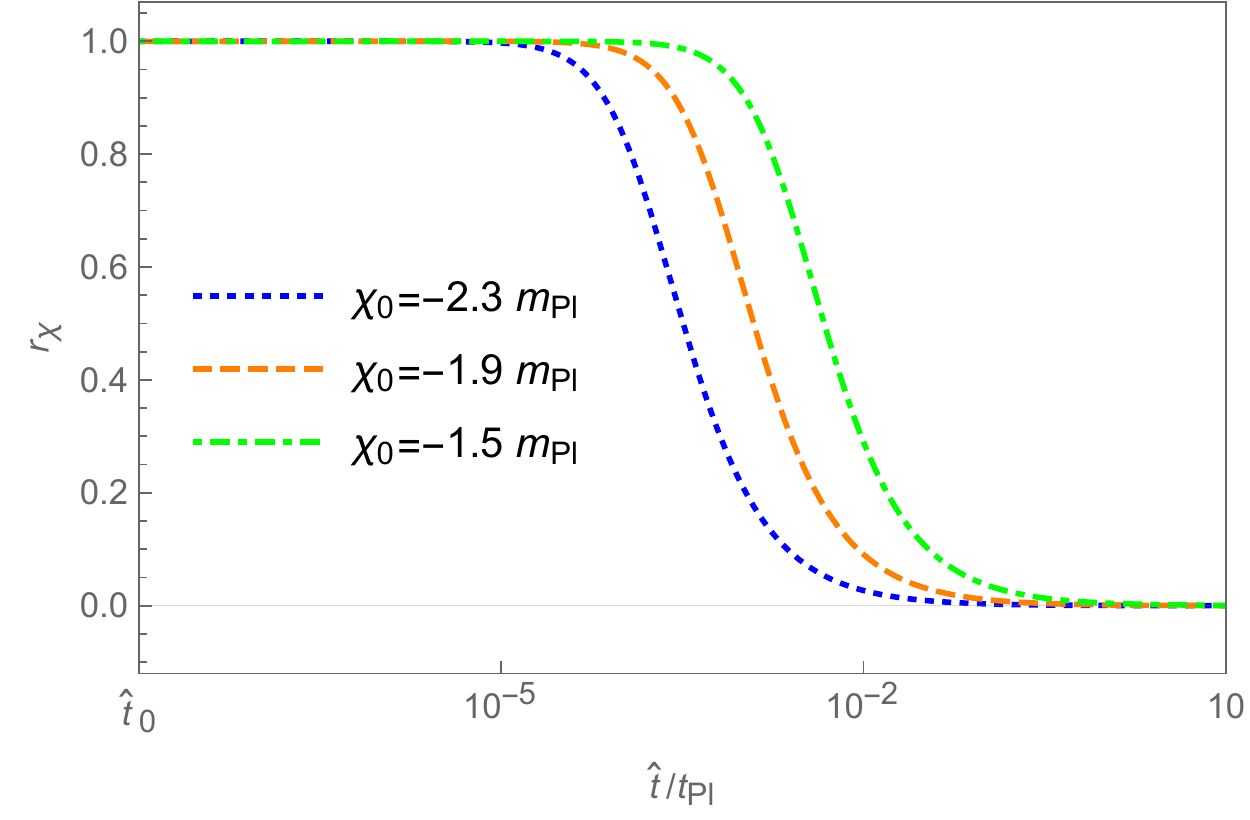}
\includegraphics[width=5.8cm]{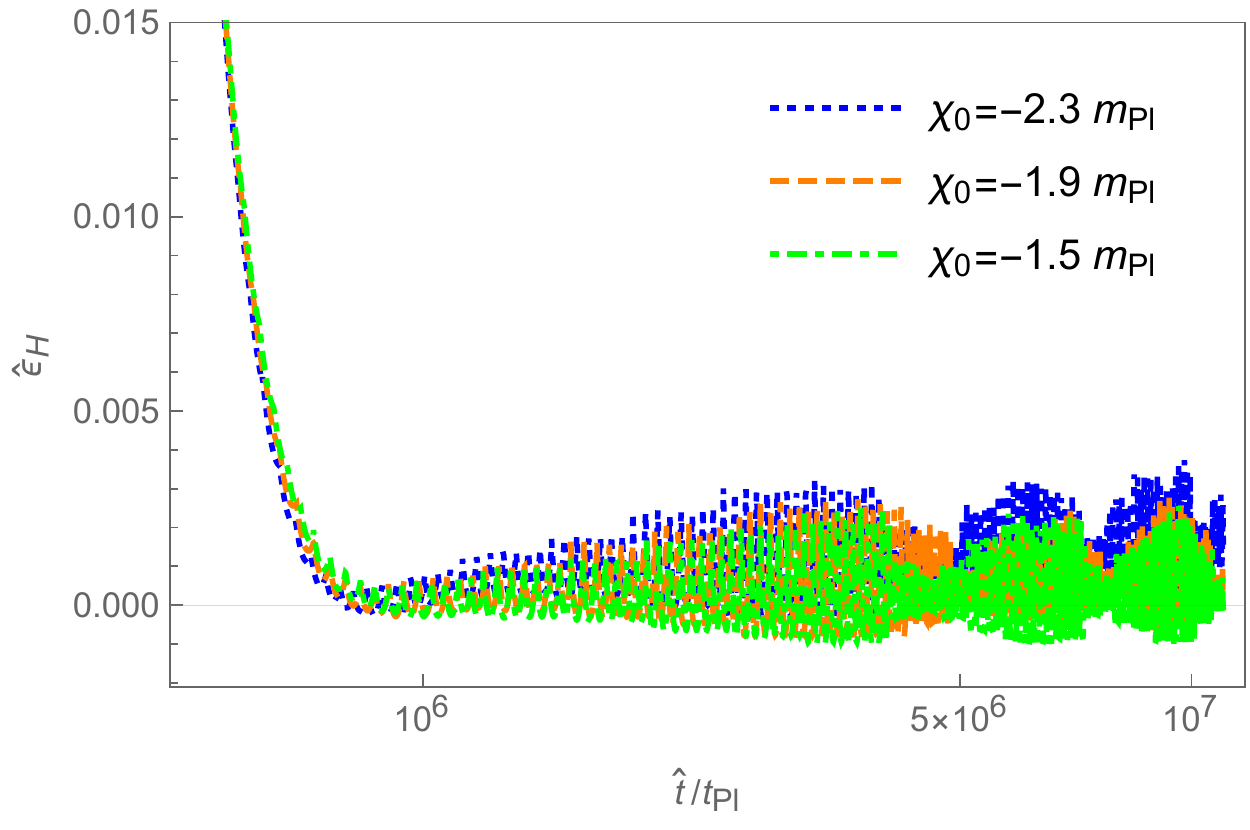}
\includegraphics[width=5.8cm]{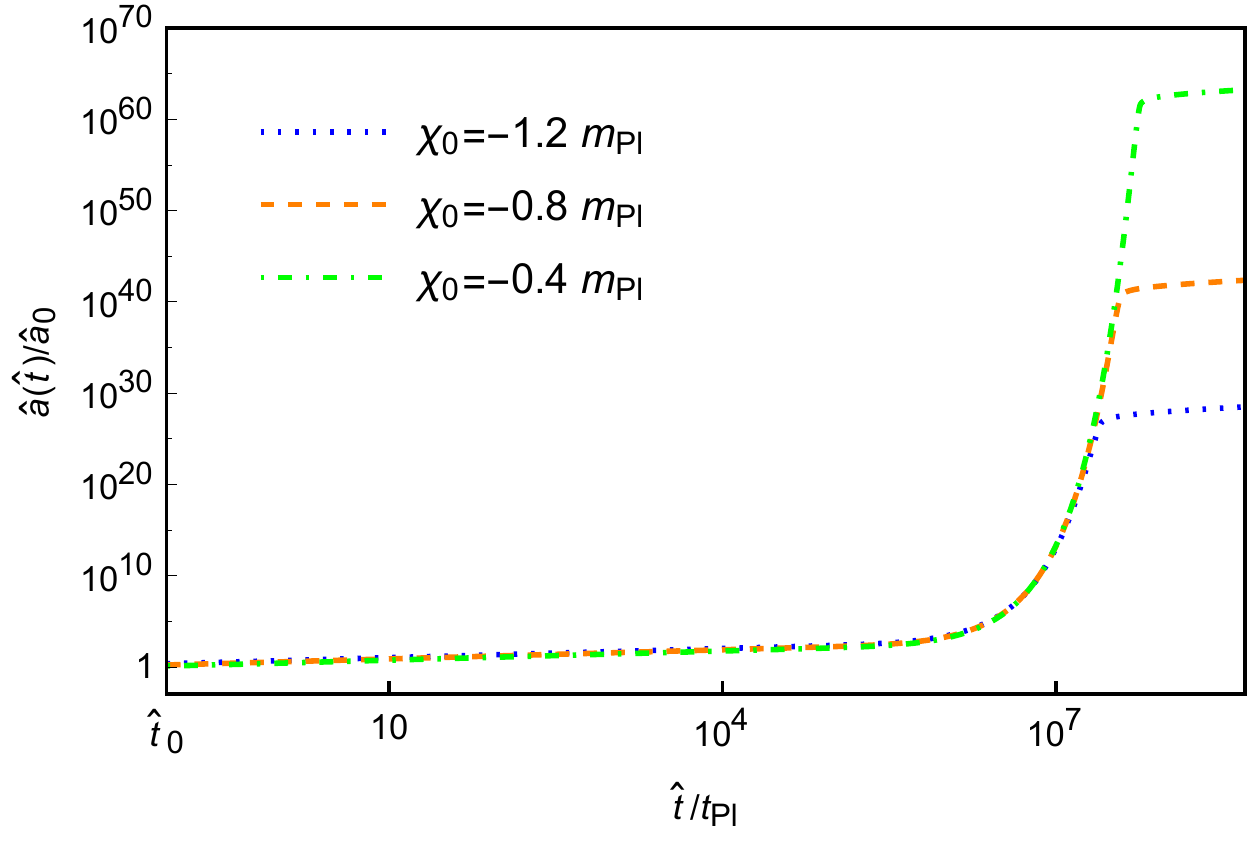}
\includegraphics[width=5.8cm]{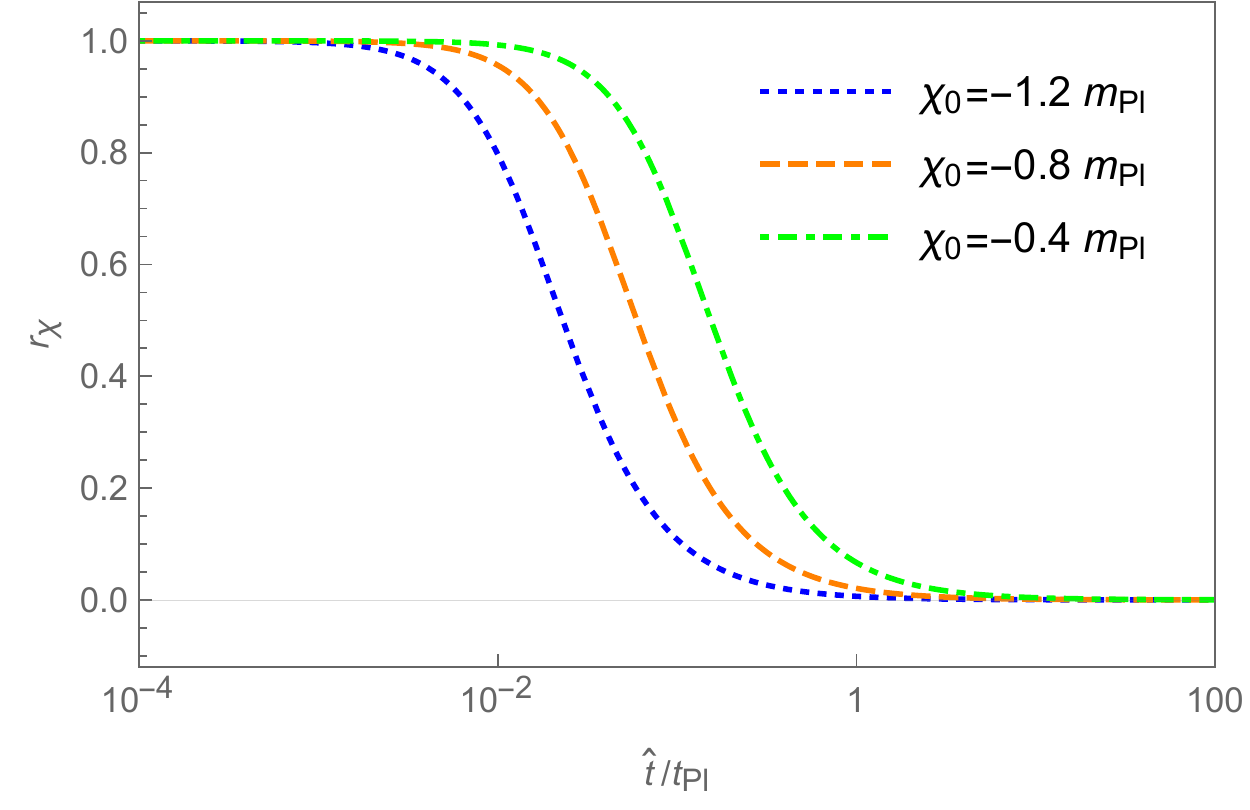}
\includegraphics[width=5.8cm]{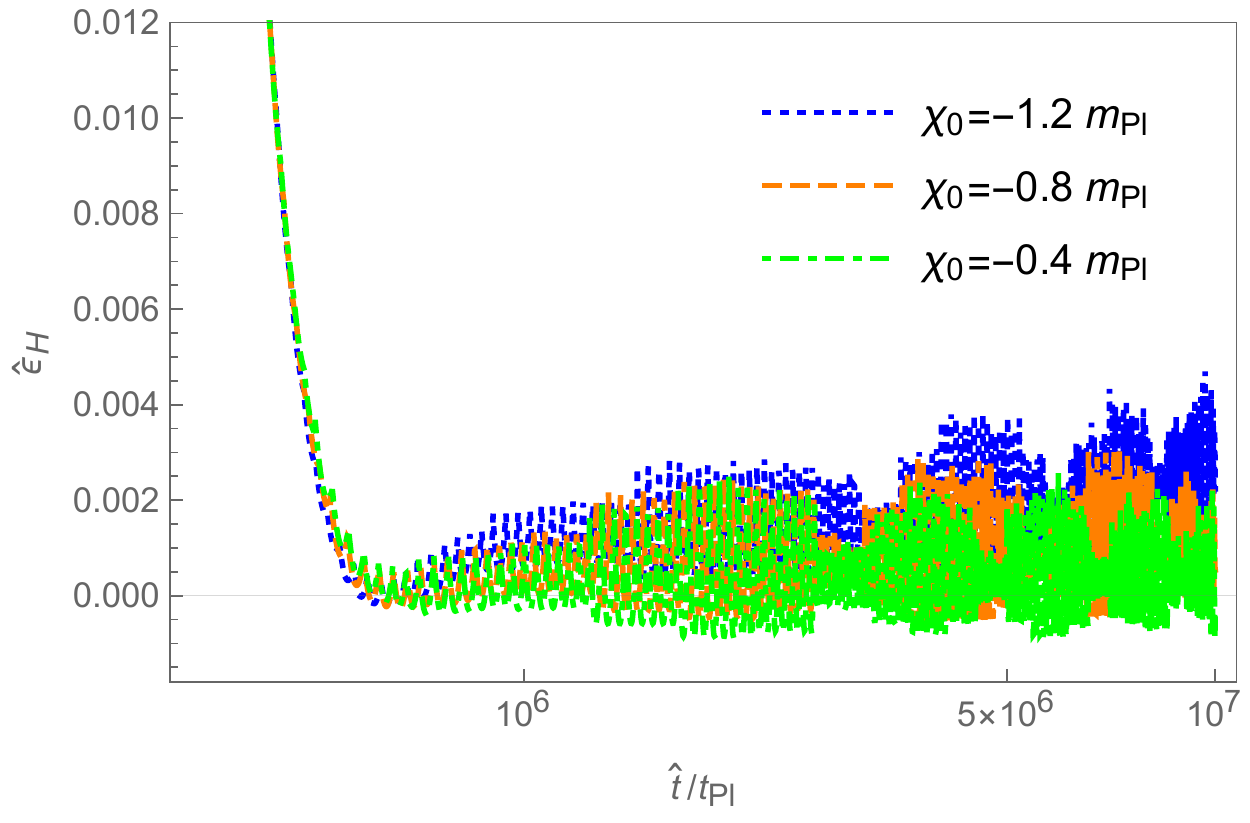}
\caption{The figure shows the numerical evolution of $\hat a(\hat t)$ ,$r_{\chi}$ and $\hat \epsilon_{H} $ with $\dot \chi_0 > 0 $. Top panel: $\beta=5$. Middle panel: $\beta=10$. Bottom panel: $\beta=20$. Contrary to the Fig.~\ref{bigfig1}, now the quantum bounce does not exist after the initial time $\hat t_0$.}
\label{bigfig2}
\end{figure*}

For $\beta=5,\;10, \;20$, the background evolutions for a set of initial conditions with $\dot \chi_0<0$ and $\dot \chi_0>0$ are illustrated, respectively, in Fig.~\ref{bigfig1} and \ref{bigfig2}. In both figures, all the three cases ($\beta=5,\;10, \;20$) are presented and the scale factor $\hat a(\hat t)$ , $r_{\chi} $ and the slow-roll parameter $\hat \epsilon_{H}$ are all obtained numerically. From these figures, similar to cases of $\beta=1$ and $\beta=3$, the evolution again can be divided into three phases, the pre-inflationary quantum phase, quantum-to-classical transition, and the slow-roll inflationary phase. For the pre-inflationary quantum phase, an important  feature is that a quantum bounce always occurs for negative initial velocity ($\dot \chi_0<0$), while it does not exist after the initial time $\hat t_0$ for positive initial velocity ($\dot \chi_0>0$) which is in contrast to cases of $\beta=1$ and $\beta=3$. The numerical results of the background evolution for various initial conditions are also presented respectively in  Table.~\ref{bigtable2} for $\dot \chi_0>0$ and Table.~\ref{bigtable1} for $\dot \chi_0 <0$.

Finally, for each cases ($\beta=5,\;10, \;20$), to obtain at least 60 $e$-folds during the slow-roll inflationary phase, the values of $\chi_{0}$ have to be restricted to the ranges given as follows: for $\beta=5$ one finds
\bqn
\chi_0 \in  
\begin{cases}
(2.12 m_{\rm Pl}, 2.95 m_{\rm Pl}), & \;  \dot \chi_{0}<0, \\
(-6.53 m_{\rm Pl},2.95 m_{\rm Pl}),  & \; \dot \chi_{0}>0,
\end{cases}
\eqn
for $\beta=10$ we have
\bqn
\chi_0 \in 
\begin{cases}
(2.55 m_{\rm Pl},4.08 m_{\rm Pl}), & \;   \dot \chi_{0}<0, \\
 (-2.18 m_{\rm Pl},4.08 m_{\rm Pl}), & \;   \dot \chi_{0}>0,
\end{cases}
\eqn
and for $\beta=20$,
\bqn
\chi_0 \in 
\begin{cases}
(3.00 m_{\rm Pl},5.65 m_{\rm Pl}), & \;   \dot \chi_{0}<0, \\
(-1.13 m_{\rm Pl},5.65 m_{\rm Pl}), & \;   \dot \chi_{0}>0.
\end{cases}
\eqn
Within the above ranges, the e-folds $N_{\rm inf}$ increases as the value of $\chi_0$ increases, as shown in Table.~\ref{bigtable2} and \ref{bigtable1}.

\begin{table*}[t]
	\centering
	\caption{Table for numerical results for $\alpha$-attractor inflation for the cases $\beta=1,\; 5,\; 10,\; 20$  with $\dot \chi_0>0$.}
	 \lb{bigtable2}
	\begin{tabular}{p{0.09\textwidth}<{\centering}p{0.11\textwidth}<{\centering}p{0.12\textwidth}<{\centering}p{0.13\textwidth}<{\centering}p{0.12\textwidth}<{\centering}p{0.13\textwidth}<{\centering}p{0.12\textwidth}<{\centering}p{0.12\textwidth}<{\centering}}
		\hline
\specialrule{0.04em}{2.2pt}{3pt} 
$\beta $          &  $\chi_0$                 &  $\hat t_{\rm B}/t_{\rm Pl}$&   $\hat t_C/t_{\rm Pl}$               & Inflation  & $\hat t/t_{\rm Pl}$        & $\chi_*$         & $N_{\rm inf}$ \\ 
\specialrule{0.04em}{3pt}{3pt}
\multirow{2}{*}{1} & \multirow{2}{*}{1.38} & \multirow{2}{*}{$ 4.1\times 10^5$} &  \multirow{2}{*}{$ 1.0\times 10^8$}  &   starts    &     $ 2.9\times 10^5$     & 1.38               & \multirow{2}{*}{472.45}  \\
          &                         &                                       &                                     &  ends      &     $ 6.9\times 10^8$     &0.10                &        \\ \specialrule{0em}{3pt}{3pt}
	     & \multirow{2}{*}{1.3}   &  \multirow{2}{*}{$ 6.0\times 10^5$}   & \multirow{2}{*}{$ 1.0\times 10^8$}   &   starts    &     $ 4.8\times 10^5$     & 1.38               & \multirow{2}{*}{472.45}  \\
          &                         &                                      &                                      &  ends      &     $ 6.9\times 10^8$     &0.10                &       \\  \specialrule{0em}{3pt}{3pt}
	    & \multirow{2}{*}{0}     &  \multirow{2}{*}{$ 7.1\times 10^5$}    &  \multirow{2}{*}{$ 1.0\times 10^8$}  &   starts    &     $ 5.9\times 10^5$     & 1.38               & \multirow{2}{*}{472.53}  \\
          &                         &                                      &                                       &  ends      &     $ 6.9\times 10^8$     &0.10                &       \\ 
\specialrule{0.04em}{3pt}{3pt}
\multirow{2}{*}{5} & \multirow{2}{*}{-5}      &   \multirow{2}{*}{None}  &  \multirow{2}{*}{$ 4.1\times 10^{-4}$}  &   starts    &     $ 2.7\times 10^5$     & 1.28               & \multirow{2}{*}{97.27}  \\
                   &                          &                       &                 &  ends      &     $ 7.0\times 10^7$     &0.13                &        \\
                    \specialrule{0em}{3pt}{3pt}
	              & \multirow{2}{*}{-5.63}   & \multirow{2}{*}{None} & \multirow{2}{*}{$ 2.1\times 10^{-5}$}   &   starts    &     $ 2.7\times 10^5$     & 1.14               & \multirow{2}{*}{60.77}  \\
                   &                          &             &                            &  ends      &     $ 4.5\times 10^7$     &0.13                &       \\  \specialrule{0em}{3pt}{3pt}
	              & \multirow{2}{*}{-7}    & \multirow{2}{*}{None}  &  \multirow{2}{*}{$ 3.1\times 10^{-8}$ } &   starts    &     $ 2.8\times 10^5$     & 0.84               & \multirow{2}{*}{21.13}  \\
                  &                           &                     &                   &  ends      &     $ 1.7\times 10^7$     &0.13                &       \\  \specialrule{0em}{3pt}{3pt}
	             & \multirow{2}{*}{-10}      & \multirow{2}{*}{None}  &  \multirow{2}{*}{$ 1.6\times 10^{-14}$}                 &   starts    &                          &                    &       \\
                  &                           &                  &                      &  ends        &                            &                    &       \\  \specialrule{0em}{3pt}{3pt}
	             & \multirow{2}{*}{-12}      & \multirow{2}{*}{None}  & \multirow{2}{*}{$ 1.4\times 10^{-18}$} &   starts    &                           &                   &      \\
                  &                           &                 &                      &  ends      &                            &                  &       \\ \specialrule{0.04em}{4pt}{4pt}
\multirow{2}{*}{10} & \multirow{2}{*}{-1.6}    & \multirow{2}{*}{None} &  \multirow{2}{*}{3.98}                 &   starts    &     $ 2.0\times 10^5$     & 1.64               & \multirow{2}{*}{113.93}  \\
                  &                            &                &                      &  ends      &     $ 6.2\times 10^7$     &0.15                &        \\ \specialrule{0em}{3pt}{3pt}
	             & \multirow{2}{*}{-2.18}    & \multirow{2}{*}{None}  & \multirow{2}{*}{0.57}                &   starts    &     $ 2.0\times 10^5$     & 1.38               & \multirow{2}{*}{60.19}  \\
                  &                            &                 &                      &  ends      &     $ 3.4\times 10^7$     &0.15                &       \\  \specialrule{0em}{3pt}{3pt}
	             & \multirow{2}{*}{-2.8}   & \multirow{2}{*}{None}    &  \multirow{2}{*}{0.07 }             &   starts    &     $ 2.1\times 10^5$     & 1.1               & \multirow{2}{*}{29.46}  \\
                  &                           &                     &                  &  ends      &     $ 1.8\times 10^7$     &0.15                &       \\  \specialrule{0em}{3pt}{3pt}
	            & \multirow{2}{*}{-5}       & \multirow{2}{*}{None}    &  \multirow{2}{*}{$ 4.3\times 10^{-5}$}              &   starts    &                          &                    &       \\
                  &                           &           &                            &  ends      &                            &                    &       \\  \specialrule{0em}{3pt}{3pt}
	            & \multirow{2}{*}{-10}        & \multirow{2}{*}{None} & \multirow{2}{*}{$ 2.1\times 10^{-12}$} &   starts    &                           &                   &      \\
                  &                           &                &                       &  ends      &                            &                  &       \\ \specialrule{0.04em}{4pt}{4pt}
\multirow{2}{*}{20}& \multirow{2}{*}{-0.5}    & \multirow{2}{*}{None}  &  \multirow{2}{*}{23.88}                &   starts    &     $ 1.5\times 10^5$     & 2.01               & \multirow{2}{*}{122.63}  \\
                 &                             &                &                       &  ends      &     $ 5.2\times 10^7$     &0.16                &        \\ \specialrule{0em}{3pt}{3pt}
	            & \multirow{2}{*}{-1.13}        & \multirow{2}{*}{None}  & \multirow{2}{*}{5.34}              &   starts    &     $ 1.5\times 10^5$     & 1.64               & \multirow{2}{*}{60.66}  \\
                 &                            &            &                            &  ends      &     $ 2.7\times 10^7$     &0.16                &       \\  \specialrule{0em}{3pt}{3pt}
	           & \multirow{2}{*}{-2}       & \multirow{2}{*}{None} &  \multirow{2}{*}{0.67 }                 &   starts    &     $ 1.7\times 10^5$     & 1.11               & \multirow{2}{*}{20.72}  \\
                 &                            &          &                             &  ends      &     $ 1.1\times 10^7$     &0.16                &       \\  \specialrule{0em}{3pt}{3pt}
	            & \multirow{2}{*}{-5}      & \multirow{2}{*}{None}   &  \multirow{2}{*}{$ 5.4\times 10^{-4}$}  &   starts    &                          &                    &       \\
                 &                            &            &                           &  ends      &                            &                    &       \\  \specialrule{0em}{3pt}{3pt}
	           & \multirow{2}{*}{-10}       & \multirow{2}{*}{None}  & \multirow{2}{*}{$ 3.7\times 10^{-9}$} &   starts    &                           &                   &      \\
                 &                             &             &                          &  ends      &                            &                  &       \\ \specialrule{0em}{3pt}{3pt}
\hline
\hline
	\end{tabular}
\end{table*}

\begin{table*}[t]
	\centering
	\caption{Table for the cases $\beta=$5, 10 and 20 with $\dot \chi_0<0$. Blank cells mean that the slow-roll inflation does not exist.} \lb{bigtable1}
	\begin{tabular}{p{0.09\textwidth}<{\centering}p{0.11\textwidth}<{\centering}p{0.12\textwidth}<{\centering}p{0.13\textwidth}<{\centering}p{0.12\textwidth}<{\centering}p{0.13\textwidth}<{\centering}p{0.12\textwidth}<{\centering}p{0.12\textwidth}<{\centering}}
		\hline
\specialrule{0.04em}{2.2pt}{3pt}
$\beta $   &  $\chi_0$   & $\hat t_B/t_{\rm Pl}$   &   $\hat t_C/t_{\rm Pl}$   & Inflation  & $\hat t/t_{\rm Pl}$        & $\chi_*$         & $N_{\rm inf}$ \\ 
\specialrule{0.04em}{3pt}{3pt}
\multirow{2}{*}{1} & \multirow{2}{*}{1.3} & \multirow{2}{*}{$ 4.5\times 10^4$} &  \multirow{2}{*}{$ 4.1\times 10^5$}  &   starts    &     $ 4.9\times 10^5$     & 1.02               & \multirow{2}{*}{161.49}  \\
          &                         &                                       &                                     &  ends      &     $ 2.4\times 10^8$     &0.10                &        \\ \specialrule{0em}{3pt}{3pt}
	     & \multirow{2}{*}{1.25}   &  \multirow{2}{*}{$ 2.6\times 10^4$}   & \multirow{2}{*}{$ 2.1\times 10^5$}   &   starts    &     $ 5.2\times 10^5$     & 0.88               & \multirow{2}{*}{61.80}  \\
          &                         &                                      &                                      &  ends      &     $ 9.3\times 10^7$     &0.10                &       \\  \specialrule{0em}{3pt}{3pt}
	    & \multirow{2}{*}{0}     &  \multirow{2}{*}{0.04}             &  \multirow{2}{*}{0.34}                    &   starts    &     $ 6.0\times 10^5$     & 0.31               & \multirow{2}{*}{4.39}  \\
          &                         &                                      &                                       &  ends      &     $ 7.9\times 10^6$     &0.10                &       \\  \specialrule{0em}{3pt}{3pt}
	     & \multirow{2}{*}{-1.81}  &  \multirow{2}{*}{$ 1.9\times 10^{-10}$} & \multirow{2}{*}{$ 1.5\times 10^{-9}$} &   starts    &  $ 5.6\times 10^5$     &  0.67                  &  \multirow{2}{*}{60.42}     \\
          &                         &                                       &                                         &  ends      &    $ 9.1\times 10^7$    &  0.10                  &       \\ 
\specialrule{0.04em}{3pt}{3pt}
\multirow{2}{*}{5}& \multirow{2}{*}{2.2}   & \multirow{2}{*}{1554.29}      &  \multirow{2}{*}{$ 3.1\times 10^4$}  &   starts    &     $ 2.6\times 10^5$     & 1.50               & \multirow{2}{*}{96.82}  \\
          &                         &                                       &                                     &  ends      &     $ 7.0\times 10^7$     &0.13                &        \\ \specialrule{0em}{3pt}{3pt}
	     & \multirow{2}{*}{2.12}   &  \multirow{2}{*}{1062.24}            & \multirow{2}{*}{$ 2.1\times 10^4$}   &   starts    &     $ 2.6\times 10^5$     & 1.35               & \multirow{2}{*}{60.13}  \\
          &                         &                                      &                                      &  ends      &     $ 4.5\times 10^7$     &0.13                &       \\  \specialrule{0em}{3pt}{3pt}
	    & \multirow{2}{*}{1.8}     &  \multirow{2}{*}{231.85}             &  \multirow{2}{*}{$ 4.6\times 10^3$ } &   starts    &     $ 2.9\times 10^5$     & 0.77               & \multirow{2}{*}{7.30}  \\
          &                         &                                      &                                       &  ends      &     $ 6.8\times 10^6$     &0.13                &       \\  \specialrule{0em}{3pt}{3pt}
	     & \multirow{2}{*}{0}      &  \multirow{2}{*}{0.04}               &  \multirow{2}{*}{0.88}              &   starts    &                          &                    &       \\
          &                         &                                       &                                     &  ends      &                            &                    &       \\  \specialrule{0em}{3pt}{3pt}
	    & \multirow{2}{*}{-2}     & \multirow{2}{*}{$3.3\times 10^{-6}$} & \multirow{2}{*}{$ 6.5\times 10^{-5}$} &   starts    &                           &                   &      \\
          &                         &                                       &                                       &  ends      &                            &                  &       \\ \specialrule{0.04em}{4pt}{4pt}
\multirow{2}{*}{10} & \multirow{2}{*}{2.7}    & \multirow{2}{*}{361.53}     &  \multirow{2}{*}{$ 1.1\times 10^4$}  &   starts    &     $ 1.9\times 10^5$     & 1.82               & \multirow{2}{*}{105.63}  \\
          &                         &                                       &                                     &  ends      &     $ 5.8\times 10^7$     &0.15                &        \\ \specialrule{0em}{3pt}{3pt}
	     & \multirow{2}{*}{2.55}   &  \multirow{2}{*}{225.77}            & \multirow{2}{*}{$ 6.9\times 10^3$}   &   starts    &     $ 2.0\times 10^5$     & 1.60               & \multirow{2}{*}{61.71}  \\
          &                         &                                      &                                      &  ends      &     $ 3.5\times 10^7$     &0.15                &       \\  \specialrule{0em}{3pt}{3pt}
	    & \multirow{2}{*}{2.4}     &  \multirow{2}{*}{131.82}             &  \multirow{2}{*}{$ 4.0\times 10^3$ } &   starts    &     $ 2.0\times 10^5$     & 1.35               & \multirow{2}{*}{32.27}  \\
          &                         &                                      &                                       &  ends      &     $ 2.0\times 10^7$     &0.15                &       \\  \specialrule{0em}{3pt}{3pt}
	     & \multirow{2}{*}{0}      &  \multirow{2}{*}{0.04}               &  \multirow{2}{*}{1.23}              &   starts    &                          &                    &       \\
          &                         &                                       &                                     &  ends      &                            &                    &       \\  \specialrule{0em}{3pt}{3pt}
	    & \multirow{2}{*}{-2}     & \multirow{2}{*}{$4.9\times 10^{-5}$} & \multirow{2}{*}{$ 1.5\times 10^{-3}$} &   starts    &                           &                   &      \\
          &                         &                                       &                                       &  ends      &                            &                  &       \\ \specialrule{0.04em}{4pt}{4pt}
\multirow{2}{*}{20}& \multirow{2}{*}{3.2}    & \multirow{2}{*}{73.05}       &  \multirow{2}{*}{$3.5\times 10^3$}  &   starts    &     $ 1.5\times 10^5$     & 2.13               & \multirow{2}{*}{103.08}  \\
          &                         &                                       &                                     &  ends      &     $ 4.4\times 10^7$     &0.16                &        \\ \specialrule{0em}{3pt}{3pt}
	     & \multirow{2}{*}{3}      &  \multirow{2}{*}{45.40}               & \multirow{2}{*}{$ 3.5\times 10^3$}   &   starts    &     $ 1.5\times 10^5$     & 1.84               & \multirow{2}{*}{61.09}  \\
          &                         &                                      &                                      &  ends      &     $ 2.8\times 10^7$     &0.16                &       \\  \specialrule{0em}{3pt}{3pt}
	    & \multirow{2}{*}{2.7}     &  \multirow{2}{*}{22.24}             &  \multirow{2}{*}{$ 1.0\times 10^3$ } &   starts    &     $ 1.6\times 10^5$     & 1.42               & \multirow{2}{*}{25.67}  \\
          &                         &                                      &                                       &  ends      &     $ 1.3\times 10^7$     &0.16                &       \\  \specialrule{0em}{3pt}{3pt}
	     & \multirow{2}{*}{0}      &  \multirow{2}{*}{0.04}               &  \multirow{2}{*}{1.75}              &   starts    &                          &                    &       \\
          &                         &                                       &                                     &  ends      &                            &                    &       \\  \specialrule{0em}{3pt}{3pt}
	    & \multirow{2}{*}{-2}     & \multirow{2}{*}{$3.1\times 10^{-4}$} & \multirow{2}{*}{$ 0.02    $}           &   starts    &                           &                   &      \\
          &                         &               &               &  ends      &                            &                  &       \\ \specialrule{0em}{3pt}{3pt}
\hline
\hline
	\end{tabular}
\end{table*}

\section{Conclusion and Discussion}
\renewcommand{\theequation}{5.\arabic{equation}}\setcounter{equation}{0}

In this paper we have provided a detailed numerical study of the Starobinsky and $\alpha$-attractor inflation as well as their pre-inflationary dynamics in the framework of loop quantum BD cosmology. We show that for both the Starobinsky and $\alpha$-attractor inflation, the evolution of the background Universe can be in general divided into three different phases: {\em pre-inflationary quantum phase, quantum-to-classical transition, and slow-roll inflation}. During the pre-inflationary quantum phase, the background evolution is dominated by the quantum geometry effects of loop quantum BD cosmology. Unlike the background evolution in LQC of GR where the pre-inflationary dynamics represents an expanding Universe started at the quantum bounce \cite{zhu_preinflationary_2017a}, the pre-inflationary dynamics is very complicated and depends on initial conditions and specific models. Generally, the Universe is initially expanding if the initial velocity of the scalar field $\chi$ is positive ($\dot \chi_0>0$) and is contracting if it is negative ($\dot \chi_0 <0$). For Starobinky inflaton($\beta=3$) and $\alpha$-attractor inflation with $\beta=1$, the initial expanding Universe shall collapse to a contracting phase before it evolves into the final expanding phase through the quantum bounce, while initial contracting Universe directly connects to the expanding phase through the quantum bounce. For $\alpha$-attractor inflation with $\beta=5, \; 10, \; 20$, we show that the quantum bounce does not exist after the initial time $\hat t_0$ for the initial expanding Universe. Whether a quantum bounce would appear before the chosen initial time $\hat t_0$ deserve further investigating. This issue concerns whether the effective equations are still valid for extremely high energy near to the classical singularity and thus is out of the scope of this paper.  For initial contracting Universe, the evolution of the background is almost the same as that in models of Starobinsky inflation ($\beta=3$) and $\alpha$-attractor inflation with $\beta=1$. After the pre-inflationary quantum phase, the universe gradually evolves into the expanding Universe. For some of initial conditions in the parameter space, we show that the slow-roll inflation for both the Starobinsky and $\alpha$-attractor models are produced. In addition, to be consistent with observational data, we also derive the range of initial conditions that could produce at least $60$ $e$-folds during the slow-roll inflation.

\section*{Acknowledgements}
This work is supported by National Natural Science Foundation of China with the Grant Nos. 11675143 (W.-J. J. \& T.Z.), 11475023 (Y.M.), and 11875006 (Y.M.).

%\bibliographystyle{unsrt}
%\bibliography{lib}

\end{document}